\documentclass{article}
\usepackage{graphicx} 
\usepackage[bottom=3cm]{geometry}
\usepackage[font=normal,labelfont=bf]{caption}
\usepackage[section]{placeins}
\usepackage{xcolor}
\usepackage{comment}
\usepackage{svg}
\usepackage[american]{babel}
\usepackage{bm}
\usepackage{booktabs}
\usepackage{multirow}
\usepackage{setspace}
\usepackage{hyperref}
\usepackage{natbib}
\usepackage [autostyle,english=american]{csquotes} 
\usepackage{authblk}
\usepackage{soul}
\usepackage{chngcntr}

\usepackage{graphicx}%
\graphicspath{{./figures/} }
\usepackage[caption=false]{subfig}

\usepackage{multirow}%
\usepackage{amsmath,amssymb,amsfonts}%
\usepackage{amsthm}%
\usepackage{mathrsfs}%
\usepackage[title]{appendix}%
\usepackage{xcolor}%
\usepackage{textcomp}%
\usepackage{manyfoot}%
\usepackage{booktabs}%
\usepackage{algorithm}%
\usepackage{algorithmicx}%
\usepackage{algpseudocode}%
\usepackage{listings}%
\usepackage{fullpage}%
\usepackage{natbib}

\usepackage{physics}
\usepackage{xltabular}
\usepackage{longtable}
\usepackage{changepage}
\usepackage{lscape}
\usepackage{geometry}
\usepackage{pdflscape}
\usepackage{caption}

\usepackage{rotating}

\definecolor{rev}{RGB}{0, 0, 0}
\usepackage{epstopdf}
\MakeOuterQuote{"} 

\title{Integrating large language models and active inference to understand eye movements in reading and dyslexia}

\author[1,\dag]{Francesco Donnarumma}
\author[1,\dag]{Mirco Frosolone}
\author[1,*]{Giovanni Pezzulo}

\affil[1]{Institute of Cognitive Sciences and Technologies, National Research Council, Rome, Italy}
\affil[$\dag$]{\textcolor{rev}{Shared first authorship}}
\affil[*]{Corresponding author: giovanni.pezzulo@istc.cnr.it}

\begin{document}

\maketitle

\abstract{We present a novel computational model employing hierarchical active inference to simulate reading and eye movements. The model characterizes linguistic processing as inference over a hierarchical generative model, facilitating predictions and inferences at various levels of granularity, from syllables to sentences. Our approach combines the strengths of large language models for realistic textual predictions and active inference for guiding eye movements to informative textual information, enabling the testing of predictions. The model exhibits proficiency in reading both known and unknown words and sentences, adhering to the distinction between lexical and nonlexical routes in dual route theories of reading. Our model therefore provides a novel approach to understand the cognitive processes underlying reading and eye movements, within a predictive processing framework. Furthermore, our model can potentially aid in understanding how maladaptive predictive processing can produce reading deficits associated with dyslexia. As a proof of concept, we show that attenuating the contribution of priors during the reading process leads to incorrect inferences and a more fragmented reading style, characterized by a greater number of shorter saccades, aligning with empirical findings regarding eye movements in dyslexic individuals. In summary, our model represents a significant advancement in comprehending the cognitive processes involved in reading and eye movements, with potential implications for understanding dyslexia in terms of maladaptive inference.}




\maketitle

\section{Introduction}\label{sec:Introduction}

Processing natural language -- encompassing understanding, reading, and producing linguistic content -- represents a fundamental ability of our species. Extensive research in psychology, neuroscience, linguistics, and machine learning has explored the intricate ways we process natural language.
Neuroscientific studies have revealed that natural language processing is inherently hierarchical, involving multiple brain regions and the integration of various sensory inputs \citep{capek2004cortical,butterfuss2018role,price2012review}. This hierarchical processing spans from individual letters, phonemes, and words to complete sentence comprehension, with a hierarchy of brain areas actively maintaining these elements in working memory across multiple time scales \citep{hasson2015hierarchical}.

Recent studies have increasingly supported the idea of \emph{hierarchical predictive coding} as a formal theory describing perception as an inferential process, involving reciprocal exchanges between predictions and prediction errors across brain hierarchies \citep{rao1999predictive,friston2005theory}. In particular, various computational neuroimaging studies employing large language models (LLMs) during linguistic tasks have provided compelling evidence for both the predictive nature of language processing and the prediction hierarchies proposed by predictive coding \citep{goldstein2022shared,schmitt2021predicting,baldassano2017discovering,zadbood2017we,chang2022information,caucheteux2023evidence,heilbron2022hierarchy}. 
A recent computational neuroimaging study investigated human electrocorticographic (ECoG) responses to narratives using LLMs trained to predict the next word. Remarkably, the study revealed that like the LLMs, the brain engages in next-word prediction before word onset, computes prediction error signals, and utilizes latent representations of words (embeddings) contextualized based on the sequence of prior words \citep{goldstein2022shared}.

Converging evidence emerges from three recent fMRI studies utilizing deep learning models during linguistic tasks. These studies not only confirm the predictive nature of language processing but also lend support to the prediction hierarchies proposed by predictive coding. The first fMRI study \citep{schmitt2021predicting} trained a LLM to predict the next word at multiple time scales, identifying event boundaries as high surprise (also explored in \citep{baldassano2017discovering,zadbood2017we,chang2022information}). Analyzing human functional magnetic resonance data during story listening, the study revealed an event-based hierarchy of surprise signals evolving along temporoparietal regions, with surprise signals gating bottom-up and top-down connectivity across neighboring time scales.
Furthermore, another fMRI study provided compelling evidence of the brain's ability to predict a hierarchy of representations spanning multiple timescales in the future \citep{caucheteux2023evidence}. Enhancing LLMs with the capability to predict beyond the next word increased their fit with human data. The study also highlighted the hierarchical organization of brain predictions, ranging from temporal cortices predicting shorter-range representations (e.g., the next word) to frontoparietal cortices predicting higher-level, longer-range, and more contextual representations.

Notably, a separate fMRI study reported that evoked brain responses to words are influenced by linguistic predictions and a metric of unexpectedness, closely aligning with the hierarchical predictive processing schemes, where lower-level predictions are informed by higher-level predictions \citep{heilbron2022hierarchy}. Additionally, the study demonstrated that these hierarchical predictions can be well-aligned with standard levels of analysis in psycholinguistics, including meaning, grammar, words, and speech sounds, reinforcing the validity of the standard decomposition.
Taken together, these studies, alongside others \citep{kutas1984brain,weissbart2020cortical,koskinen2020brain,willems2016prediction,frank2015erp}, provide compelling support for the significance of prediction and hierarchical predictive coding in language processing. Moreover, they underscore the growing relevance of LLMs in comprehending linguistic processing \citep{goldstein2022shared}.

Despite these advancements, the above studies have primarily focused on LLMs that passively receive sensory information, rather than actively searching for it. However, linguistic tasks, such as reading written text and listening to speech, are inherently active processes \citep{ferro2010reading,norris2006bayesian,friston2021active}. For example, during reading, eye movements (saccades) actively guide attention to relevant parts of the text, rather than processing every piece of text linearly. This active reading process suggests that saccades play a crucial role in hypothesis testing, selecting informative parts of the text to test predictions \citep{ferro2010reading,norris2006bayesian,donnarumma2017action,friston2012perceptions}.



Several reading models incorporating eye movements, such as E-Z Reader \citep{reichle1998toward}, SWIFT \citep{engbert2005swift,seelig2020bayesian,rabe2021bayesian}, Über-Reader \citep{reichle2021computational,veldre2020towards}, Glenmore \citep{reilly2002glenmore,reilly2006some}, SEAM \citep{rabe2024seam}, the rational model of eye movements \citep{bicknell2010rational} and OB1-Reader \citep{snell2018ob1} significantly advanced our understanding of the role of reading dynamics. These models have successfully replicated numerous empirical findings, highlighting how word-level attributes like length, frequency, and predictability impact reading dynamics. However, these models do not exploit the generative capabilities of recent large language models and do not fully align with the aforementioned evidence of hierarchical predictive processing supporting reading and eye movements.

In this paper, we propose a novel computational model that unifies hierarchical predictive processing and hypothesis testing during reading by integrating the LLM BERT (Bidirectional Encoder Representations from Transformers) \citep{Devlin20194171} with active inference \citep{friston2010free,parr2022active}: a theory of perception and action based on Bayesian principles, whereby agents minimize expected surprise (or variational free energy) through iterative belief updating and action selection. Our model views reading as an active (Bayesian) inference problem, employing a hierarchical generative model to represent causal relationships between textual elements at different levels (letters, syllables, words, and sentences). By generating predictions at each level and testing them through saccades, our model actively simulates reading.

The model incorporates three significant insights. Firstly, it conceptualizes linguistic processing as inference, employing a hierarchical generative model that allows for predicting and inferring at different timescales, such as syllables, words, and sentences. By integrating the LLM BERT \citep{Devlin20194171} at the highest hierarchical level, our model can handle realistic reading tasks effectively, processing sentences of arbitrary length. Additionally, thanks to its hierarchical structure, our model can read both known words word-by-word and unknown words syllable-by-syllable, akin to lexical and nonlexical routes in dual route theories of reading  \citep{coltheart2005modeling}.

Secondly, the model utilizes its generative capabilities not only for language recognition and prediction, similar to LLMs, but also for simulating eye movements and saccades. In our model, reading involves an active, hypothesis testing process: the model generates saccades to the most informative parts of the text to validate its predictions and disambiguate among competing hypotheses about the content being read  \citep{ferro2010reading, baldassano2017discovering,donnarumma2017action,donnarumma2017you,friston2012perceptions,norris2006bayesian}.

Thirdly, our model can be potentially used to gain insights on reading deficits, such as dyslexia. Dyslexia, a common reading disorder affecting 5-10\% of the population, is associated not only with atypical brain activation patterns during language processing \citep{price2012review} but also with atypical eye movement patterns, such as increased numbers of forward saccades and decreased saccade lengths compared to control groups \citep{zoccolotti2005word,de1999eye,lyon2003definition,franzen2021individuals,hutzler2004eye,rayner1983eye}. Crucially, it has been suggested that dyslexia could be characterized as a disorder of inference: a computational difficulty in effectively combining prior information, such as implicit memory of previous words, with noisy observations like the currently perceived word \citep{jaffe2015computational}. Specifically, the proposal is that the reading difficulties might arise because prior information is assigned excessively low weighting (or precision) relative to its internal noise. The low precision-weighting increases the time required to form a coherent understanding of text (or to perform auditory discrimination, as in \citep{jaffe2015computational}), because novel observations consistently surprise dyslexics. By incorporating this proposal in our model, we provide a proof of concept that it can qualitatively reproduce eye movement patterns associated with dyslexia.



In the following sections, we demonstrate the model's capabilities through simulations of word reading (\emph{Simulation~1}), sentence reading (\emph{Simulation~2}), reading unknown words and sentences (\emph{Simulation~3}), and reading with prior information about the topic (\emph{Simulation~4}).

\section{Methods}
\label{sec:methods}

In this section, we provide a summary description of the model, a brief introduction to active inference \citep{parr2022active}, and a detailed discussion of the hierarchical active inference reading model used in this paper. 

\subsection{Brief explanation of the model}

\begin{figure}[ht!]
\centering
\vspace{-5em}
    \includegraphics[width=0.7\linewidth] {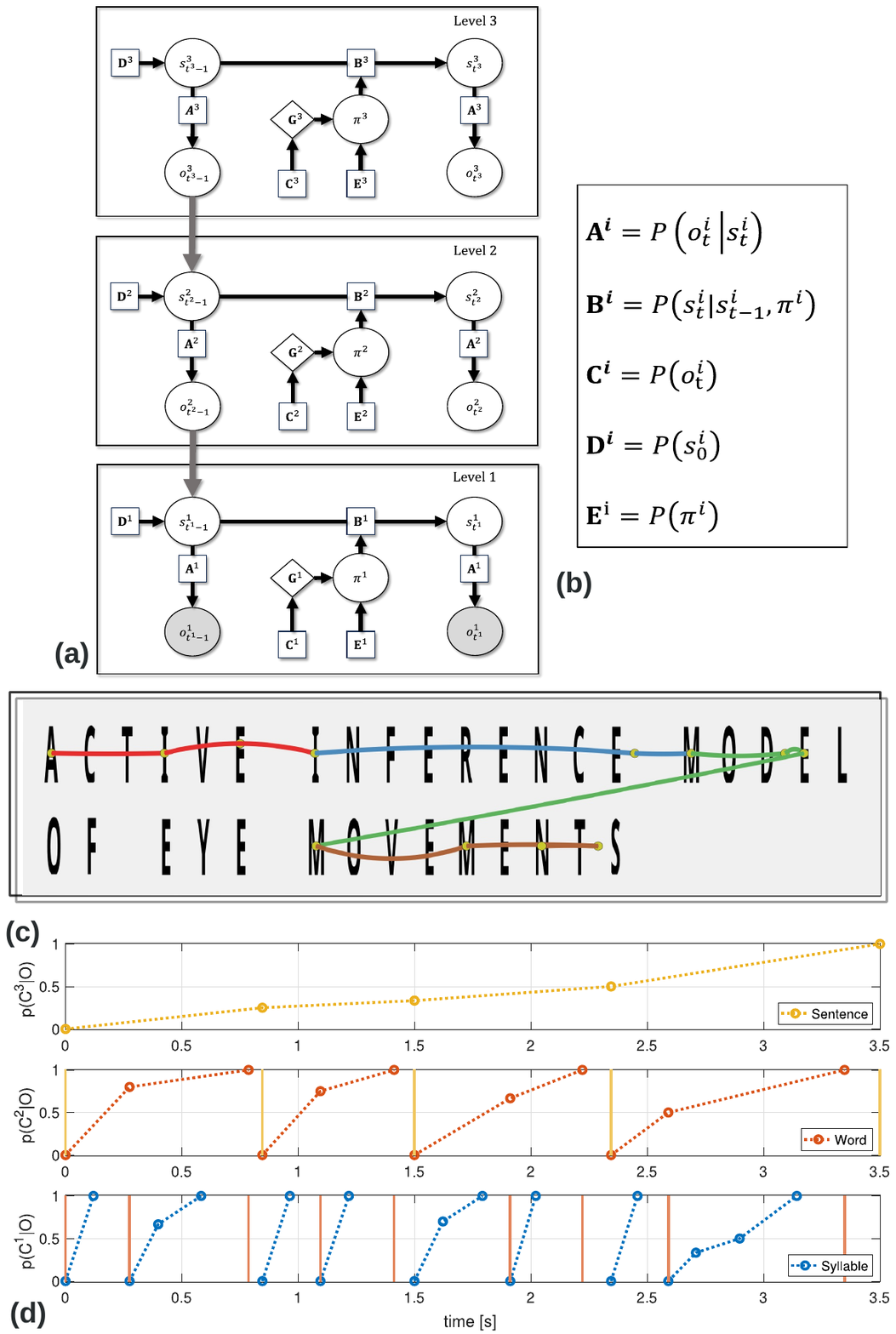}
    \caption{The hierarchical active inference model of reading and eye movements. (a) The hierarchical model comprises three layers for syllables (Level 1), words (Level 2), and sentences (Level 3), represented using the formalism of Partially Observable Markov Decision Processes (POMDPs). Empty nodes indicate hidden variables, filled nodes represent observations, and edges depict probabilistic relations between the nodes, as detailed in B. (b) Model parameters. (c) The sequence of saccades generated by the model while reading an example sentence. Yellow dots indicate the positions of the saccades in the text, and colored lines trace the sequence of saccades from red to yellow. (d) Evolution of the probabilities of correctly recognized syllables (blue lines), words (orange lines), and sentences (yellow lines) over time, while reading the sentence of Panel c. The blue circles at the bottom level indicate saccades. Once the first level confidently infers a syllable (or when the maximum level of iterations is reached), it sends a bottom-up signal to the second level (indicated by the orange vertical bar) to aid in word inference. Similarly, the second level sends a signal to the third level (indicated by the yellow vertical bar) upon confidently inferring a word. Please refer to the main text for further explanation.}
    \label{fig:HAI_model}
\end{figure}

We present a novel hierarchical active inference model for reading and eye movements during reading \citep{friston2010free}. Consistent with evidence of hierarchical language processing in the brain \citep{heilbron2022hierarchy}, our model consists of three levels representing syllables, words, and sentences, as illustrated in Figure~\ref{fig:HAI_model}(a-b).
At each level, the inference process involves integrating three types of messages: bottom-up messages from nodes at the level below (conveying observations), top-down messages from nodes at the level above (conveying predictions), and lateral messages from nodes representing the previous timestep at the same level (providing memory). For instance, the inference about the current syllable is informed by bottom-up observations about the currently observed letter, top-down predictions from the currently inferred word, and lateral information about the previous syllable. Similarly, the inference about the current word relies on bottom-up observations (i.e., the currently estimated \textcolor{rev}{syllable}), top-down predictions from the level above (i.e., at the sentence level), and \textcolor{rev}{lateral} information about the previous word.

\pagestyle{empty}

\textcolor{rev}{To ensure that our model is capable of reading realistic text, we incorporate the large language model BERT \citep{Devlin20194171} to generate prior probability distributions over the current word when reading single words, and over the next words based on previously inferred words when reading sentences. BERT was chosen because it offers computationally efficient access to explicit probability distributions over upcoming words and sentence continuations. Given an initial context—a sequence of already read words—and a predictive horizon of $H$ steps, BERT produces $K$ candidate sentence completions. These candidates are used to construct nested multinomial distributions over sentences, their constituent words, and the syllables comprising those words. These distributions then serve as priors for the three levels of the hierarchical active inference model, thereby initializing the inference process to estimate the most plausible continuation of the sentence. Inference proceeds over successive time steps as new words are observed. After a fixed number of inference steps, BERT is queried again with the updated context—including the newly read words—and the process is repeated. This iterative procedure enables the model to read and infer sentences of varying lengths.}

Lastly, our model has the capability to determine where to make (forward or backward) saccades and select the next letter observation \citep{donnarumma2017action}. The process of selecting where to direct the next saccade (i.e., which position in the syllable) is guided by the expected information gain of lexical elements across all three levels of the model. Specifically, saccades are directed to the syllable position where the resulting letter observation is expected to reduce uncertainty about the current syllable (Level 1), word (Level 2), and sentence (Level 3); see Section \ref{sec:methods} for further details.

Figure \ref{fig:HAI_model}(c) visually represents the saccades generated by our model while reading an example sentence: \emph{Active inference model of eye movements}. The model demonstrates uncertainty about the first word (\emph{active}) and therefore reads it using two saccades to the first (\emph{a}) and fourth (\emph{i}) letters. It follows the same process for the subsequent two words (\emph{inference} and \emph{model}). However, at this stage, it skips the next two words, as it perceives the most informative word to be the one following (\emph{movements}). This mechanism showcases the model's ability to dynamically select the most relevant information during the reading process.

Figure \ref{fig:HAI_model}(d) illustrates the probabilities of correctly recognized syllables (blue), words (red), and sentences (yellow) during the reading process. Over time, these probabilities approach one, indicating successful recognition. Additionally, the figure demonstrates that the levels operate at different time scales, with lower levels processing information more quickly than higher levels. This temporal separation arises because each level sends messages to the level above only after accumulating sufficient confidence (as shown by the vertical bars in the figure), a process that typically requires multiple rounds of inference  \citep{friston2017active}. Note that the time it takes to arrive at a high probability of syllable can be interpreted as saccade duration -- implying that the longer the time required to resolve uncertainty about the fixated syllable, the longer the saccade duration.

\subsection{Brief introduction to Active Inference}

Active Inference is a framework that models an agent's action-perception loop by minimizing variational free energy \citep{parr2022active}. The key objective is to minimize free energy, and to achieve this, the agent possesses a generative model (like the one illustrated in Figure~\ref{fig:HAI_model}), which captures the joint probability of the stochastic variables (hidden states and observations) using the formalism of probabilistic graphical models \citep{bishop2006pattern}. The agent's perception involves making inferences about the hidden states based on the observed sensory inputs, while action involves selecting actions that can change the hidden states and optimize the model's predictions. The generative model for active inference is defined as follows:
\begin{align}
& 
\hspace{-1em}
P\left(o_{0:K},s_{0:K},u_{1:K},\gamma|\boldsymbol{\Theta}\right)=\nonumber \\
& 
\hspace{-1em}
P\left(\gamma|\boldsymbol{\Theta}\right)
P\left(\pi|\gamma,\boldsymbol{\Theta}\right)
P\left(s_{0}|\boldsymbol{\Theta}\right)
\prod_{t=0}^{K}
P\left(o_{t}|s_{t},\boldsymbol{\Theta}\right)
P\left(s_{t+1}|s_{t},\pi_{t},\boldsymbol{\Theta}\right)
\label{eq:area}
\end{align}
where 
\begin{itemize}
    \item $P\left(o_{t}|s_{t},\boldsymbol{\Theta}\right)=\mathbf{A}$, 
    \item ${P\left(s_{t+1}|s_t,\pi_t,\boldsymbol{\Theta}\right)=\mathbf{B}\left(u_t=\pi_t\right)}$, 
    \item ${P\left(\pi_t|\gamma,\boldsymbol{\Theta}\right)=\sigma\left(\ln\mathbf{E}-\gamma\cdot\mathbf{G}^{\boldsymbol{\Theta}}\left(\pi_t\right)\right)}$, 
    \item $P\left(\gamma,\boldsymbol{\Theta}\right)\sim\Gamma\left(\alpha,\beta\right)$, and

    \item $P\left(s_0|\boldsymbol{\Theta}\right)=\mathbf{D}$.

\end{itemize}

The set $\boldsymbol{\Theta}=\{\mathbf{A},\mathbf{B},\mathbf{C},\mathbf{D},\mathbf{E},\alpha,\beta \}$ parametrizes the generative model:
\begin{itemize}
    \item The (likelihood) matrix $\mathbf{A}$ encodes the relations between the observations $\mathrm{O}$ and the hidden causes of observations $\mathrm{S}$.
    \item The (transition) matrix $\mathbf{B}$ defines how hidden states evolve over time $t$, as a function of a control state (action) $u_t$; note that a sequence of control states $u_1$,$u_2$,$\dots$,$u_t$,$\dots$ defines an \emph{action policy} (or \emph{policy} for short) $\pi_t$. 
    \item \textcolor{rev}{The matrix $\mathbf{C}$ encodes an a-priori probability distribution over observations $P\left(o_\tau \right)$ and typically encodes the agent's preferences; this prior distribution parameter is needed for the resolution of Equation \eqref{eq:EFE}, with $P\left(o_\tau \right)=\mathbf{C}$.}
    \item The matrix $\mathbf{D}$ is the prior belief about the initial hidden state, before receiving any observation. 
    \item $\mathbf{E}$ encodes a prior over the policies (reflecting habitual components of action selection).
    \item $\gamma\in\mathbb{R}$ is a \emph{precision} that regulates action selection and is sampled from a $\Gamma$ distribution, with parameters $\alpha$ and $\beta$. 
\end{itemize}

In active inference, the process of perception involves estimating hidden states based on observations and previous hidden states. At the start of the simulation, the model has access to an initial state estimate $s_0$ through $\mathbf{D}$ and receives an observation $o_0$ that helps refine the estimate by using the likelihood matrix $\mathbf{A}$. Subsequently, for each time step $t=1,\dots,K$, the model infers its current hidden state $s_t$ by considering the transitions determined by the control state $u_t$, as specified in $\mathbf{B}$. Active inference employs an approximate posterior over (past, present, and future) hidden states and parameters ($s_{0:K},u_{1:K},\gamma$). This approach utilizes a factorized form of variational inference, which corresponds to a framework developed in physics known as mean field theory \citep{Parisi:111935}. Using a mean field approximation, namely assuming that all variables are independent, the factorized approximated posterior can be expressed as:
\begin{equation}
Q\left(s_{0:K},u_{1:K},\gamma\right)=Q\left(\pi\right)Q\left(\gamma\right)\prod_{t=0}^{K}Q\left(s_t|\pi_t\right)
\label{eq:mf}
\end{equation}
where the sufficient statistics are encoded by the expectations $\boldsymbol{\mu}=\left(\mathbf{\tilde{s}}^{\boldsymbol{\pi}},\boldsymbol{\pi},\boldsymbol{\gamma}\right)$, 
with $\mathbf{\tilde{s}}^{\boldsymbol{\pi}}
=\mathbf{\tilde{s}}_0^{\boldsymbol{\pi}},
\dots,\mathbf{\tilde{s}}_K^{\boldsymbol{\pi}}$. 
\color{rev}
Following a variational approach, the approximate posterior $Q(s_t|\pi_t)$ over hidden states under a given policy $\pi_t$ at each time step $t$ is optimized by minimizing the \textit{variational free energy ($\mathbf{F})$} (see \citep{parr2022active}) defined as:
\begin{equation}
\mathbf{F}_t = D_{KL}[ Q(s_t|\pi_t) \parallel P(s_t|\pi_t) ] - \mathbb{E}_{Q(s_t|\pi_t)}[\ln P(o_t | s_t)]
\label{eq:VFE}
\end{equation}

The first term, $D_{\mathrm{KL}}\left[ Q(s_t|\pi_t) \,\|\, P(s_t|\pi_t) \right]$, is the \emph{Kullback-Leibler (KL) divergence} between the approximate posterior and the prior over hidden states. This term quantifies the \emph{complexity} of the posterior—i.e., how much information (in bits or nats) is required to move from prior beliefs $P(s_t|\pi_t)$ to the posterior $Q(s_t|\pi_t)$. Minimizing this term encourages the posterior belief to have low complexity, i.e., to remain close to prior expectations.

The second term, $-\mathbb{E}_{Q(s_t|\pi_t)}\left[\ln P(o_t | s_t)\right]$, corresponds to the expected negative log-likelihood (or \emph{inaccuracy}) of the observation $o_t$ given the inferred hidden states. Minimizing this term encourages the posterior beliefs to be predictive of the actual sensory observations, thus promoting \emph{accuracy}. In sum, minimizing the variational free energy trades off complexity against accuracy. 

The approximate posterior $Q(s_t|\pi_t)$ is optimized when the sufficient statistics are:

\color{black}
\begin{subequations}
\begin{align}
\boldsymbol{s}_{t}^{\boldsymbol{\pi}}  & \approx  \sigma\left(\ln\mathbf{A}\cdot o_{t}+\ln\left(\mathbf{B}\left(\pi_{t-1}\right)\cdot\boldsymbol{s}_{t-1}^{\boldsymbol{\pi}}\right)\right)
\label{eq:suffs}\\
\boldsymbol{\pi}  & = \sigma\left(\ln\mathbf{E}-\boldsymbol{\gamma}\cdot\mathbf{G}\left(\pi_{t}\right)\right)
\label{eq:suffpi}\\
\boldsymbol{\gamma} & = \frac{\alpha}{\beta-\boldsymbol{G}\left(\boldsymbol{\pi}\right)}
\label{eq:suffg} 
\end{align}
\end{subequations}

Action selection in the active inference framework involves choosing a policy, represented by a sequence of control states $u_1,u_2,\dots,u_t$, that is expected to minimize free energy most effectively in the future. The policy distribution $\boldsymbol{\pi}$ is defined by the Softmax function $\sigma(\cdot)$. 
\textcolor{rev}{In this equation, a crucial rule is played by the \emph{expected free energy (EFE)} of the policies, denoted by $\mathbf{G}$. The EFE incorporates goal-directed components of action selection and can be interpreted as the expected variational free energy under future outcomes.}
Additionally, in this equation the precision term $\gamma$ plays a role in encoding the confidence of beliefs concerning $\mathbf{G}$. 

The EFE $\mathbf{G}(\pi_t)$ of each policy $\pi_t$ is defined as:
\begin{equation}
\hspace{-1em}
\mathbf{G}\left(\pi_{t}\right) = \sum_{\tau=t+1}^{K}D_{KL}\left[Q\left(o_{\tau}|\pi\right)\parallel P\left(o_{\tau}\right)\right] +\mathbb{E}_{\tilde{Q}}\left[H\left[P\left(o_{\tau}|s_{\tau}\right)\right]\right]
\label{eq:EFE}
\end{equation}
where $D_{KL}\left[\cdot\parallel\cdot\right]$ and $H\left[\cdot\right]$ are, respectively, the Kullback-Leibler divergence and the Shannon entropy,
$Q\left(o_{\tau},s_{\tau}|\pi\right)\triangleq P\left(o_{\tau},s_{\tau}\right)Q\left(s_{\tau}|\pi\right)$ 
is the predicted posterior distribution, $Q\left(o_{\tau}|\pi\right)=\sum_{s_{\tau}}Q\left(o_{\tau},s_{\tau}|\pi\right)$  is the predicted outcome, $P\left( o_\tau \right)$ is a categorical distribution representing the preferred outcome and encoded by $\mathbf{C}$, and $P\left(o_\tau|s_\tau\right)$ is the likelihood of the generative model encoded by the matrix $\mathbf{A}$. 

The EFE in \eqref{eq:EFE} can be used as a quality score for the policies and has two terms:

\begin{itemize}
    \item \emph{Expected Cost}. The first term of \eqref{eq:EFE} is the Kullback-Leibler divergence between the (approximate) posterior and prior over the outcomes and it constitutes the \emph{pragmatic} (or utility-maximizing) component of the quality score. This term favours the policies that entail low risk and minimise the difference between predicted ($Q\left(o_\tau|\pi\right)$) and preferred ($P\left(o_\tau \right)\equiv\mathbf{C}$) future outcomes. 
    \item \emph{Expected Ambiguity}. The second term of \eqref{eq:EFE} is the expected entropy under the posterior over hidden states and it represents the \emph{epistemic} (or uncertainty-minimizing) component of the quality score. This term favours policies that lead to states that maximize \emph{information gain} and diminish the uncertainty of future outcomes $H\left[P\left( o_\tau|s_\tau \right) \right]$.

\end{itemize}

After scoring all the policies using the EFE, action selection is performed by drawing over the action posterior expectations derived from the sufficient statistic $\boldsymbol{\pi}$ computed via \eqref{eq:suffpi}. Then, the selected action is executed, the model receives a novel observation and the perception-action cycle starts again. We refer the reader to \citep{parr2022active} for a complete derivation of the equations presented in this section.

\subsection{Hierarchical Active Inference model for reading} 

The active inference model employed in this study is structured hierarchically, consisting of three levels denoted by $i\in{1,2,3}$ in our simulations. At each level, the model encodes hidden variables associated with different aspects of textual content: syllables at Level 1, words at Level 2, and sentences at Level 3. This hierarchical organization allows the model to effectively capture and process information at multiple linguistic scales, enabling the inference and generation of syllables, words, and sentences during the reading process. For a visual representation of the model's architecture, refer to Figure~\ref{fig:HAI_model}(a-b).

The hidden states of each level are obtained by the tensorial product: $S^{(i)}=S^{(i)}_1 \otimes S^{(i)}_2 \otimes S^{(i)}_3$ among three factors: 

\begin{itemize}
    \item the \emph{Content} $S^{(i)}_1$, i.e., the textual content proper: \emph{Syllable}, \emph{Word} or \emph{Sentence},
    \item the \emph{Location} $S^{(i)}_2$, i.e. the position of a portion of the textual content (Location of the Letter in the syllable, Location of the Syllable in the word, and Location of the Word in the Sentence), thus corresponding to a $S^{(i-1)}_1$-type content of the next lower level if $i>1$ or to location of a \emph{Letter} in a \emph{Syllable} if $i=1$.
    \item the \emph{Topic} $S^{(i)}_3$, i.e. the context to which each \emph{Content} variable is associated. Please note that we only use this factor in Simulation 4.
\end{itemize}  

The Content is the crucial element of the chain: it is the text that is obtained from the recursive concatenation of the elements from the level below. A \emph{Syllable} is a concatenation of $N^{(1)}$ \emph{Letters} $c^{(0)}$ (which are not hidden states but observations, belonging to a finite \emph{Alphabet} $S^{(0)}_1$, see below). A \emph{Word} is a concatenation of $N^{(2)}$ syllables $c^{(1)}$ belonging to the set of \emph{Syllables} $S^{(1)}_1$ and a \emph{Sentence} is concatenation of $N^{(3)}$ \emph{Words} $c^{(2)}$ belonging to the set $S^{(2)}_1$.
In general, the $i$-th content $c^{(i)} \in S^{(i)}_1$ is a concatenation of $N^{(i)}$ sub-contents $c^{(i-1)}_n \in S^{(i-1)}_1$ such that  
$$ c^{(i)} = c^{(i-1)}_1, \dots c^{(i-1)}_{N^{(i)}}$$

The observations $O^{(i)}=O^{(i)}_1\otimes O^{(i)}_2 \otimes O^{(i)}_3$ consists of the tensorial product among:
\begin{itemize}
\item{the observation $O^{(i)}_1$ of the content of the level below $S^{(i-1)}_1$. Note that only Level 1 of the model receives an actual textual observation: namely, a letter  $c^{(0)}$ belonging to the Alphabet $S^{(0)}_1$. The other two levels receive as observations the content at the level below: the observations at levels 2 and 3 are the inferred syllables and words, respectively;}
\item{the observation  $O^{(i)}_2$ corresponding to the location $S^{(i)}_2$ of the $S^{(i-1)}_1$ content;}
\item{the feedback response $r \in O^{(i)}_3$}, (``correct" or ``wrong") that reports whether the currently inferred content $C^{(i)}$ has been correctly classified (i.e., assigned to the correct topic) or not. We only use this observation at Level 3 (i.e., $ O^{3}_3$) of Simulation 4, in order to compare models with or without prior information about the topic of the sentence. 
\end{itemize}

The likelihood mapping $p(O^{(i)}|S^{(i)})$ between hidden states $S^{(i)}$ and observations $O^{(i)}$ is specified through the tensor $\mathbf{A}^{(i)}$, defined as the tensorial product $\mathbf{A}^{(i)}=\mathbf{A}^{(i)}_1\otimes \mathbf{A}^{(i)}_2\otimes \mathbf{A}^{(i)}_3$ where
\begin{itemize}
\item{$\mathbf{A}^{(i)}_1$ is a 4-order tensor mapping the hidden states $S^{(i)}$} (3-order tensors) to the observation $O^{(i)}_1$ corresponding to the content  $S^{(i-1)}_1$ if $i>1$ or to the observation of a letter if $i=1$; In our simulations, this probability is typically an identity matrix: the probability of observing content  $c^{(i)}\in S^{(i)}_1$ if the reader agent is in the corresponding location $l^{(i)}\in S^{(i)}_2$ is set to 1. Note that adding noise to the likelihood function would allow modeling the effects of perceptual deficits or poor familiarity with words.
\item{$\mathbf{A}^{(i)}_2$ is a 4-order tensor mapping the hidden states $S^{(i)}$} to observation $O^{(i)}_2$ corresponding to the locations $S^{(i)}_2$; Also in this case we assume no noise, so the probability of observing a given location when the reader agent eye point to that location is set to 1.
\item{$\mathbf{A}^{(i)}_3$ is a 4-order tensor mapping the hidden states $S^{(i)}$} to the feedback response in $O^{(i)}_3$. In our simulation we adopt a binary probability in our simulation: the likelihood is set to 1 if the content belongs to the correct topic and to 0 otherwise. Please note that in this paper this factor is used only in Simulation 4.
\end{itemize}

The mapping $p ( s^{(i)}|s^{(i)}_{t-1},\pi_t )$ between hidden states given the control state u is specified by the tensor $\mathbf{B}^{(i)}$, defined as the tensorial product  $\mathbf{B}^{(i)}=\mathbf{B}^{(i)}_1\otimes \mathbf{B}^{(i)}_2\otimes \mathbf{B}^{(i)}_3$ where
\begin{itemize}
    \item \textcolor{rev}{$\mathbf{B}^{(i)}_1$ maps the transition between content states at successive time steps. Note that in our simulations, we use transition models without noise (for the \emph{control model}) and with noise added at one or more hierarchical levels (for the \emph{dyslexic models}). For this, we formalize the mapping $p( c^{(i)}_t|c^{(i)}_{t-1},\pi_t)$ by assigning $1-\delta$ to the correct target transition and assigning the value $\delta / ( M^{(i)}-1)$ across the other $M^{(i)}-1$ states, where $M^{(i)}$ is the total number of Content states $S_1^{(i)}$. This modulating factor $\delta^{(i)}\in[0,1]$, is set to 0 in the \emph{control model} and to 0.15 in the \emph{dyslexic models}.}
    \item $\mathbf{B}^{(i)}_2$ maps the transitions between locations. We equip the model with the probability $p(l^{(i)}_t|l^{(i)}_{t-1},\pi_t)$ to jump from one location $l^{(i)}\in S^{(i)}_2$ to another, without noise (i.e., the act of jumping to the selected location is executed with no errors). While in principle one can jump to any location, here we restrict the jump possibilities, by setting priors over locations $\mathbf{D}^{(i)}$ and policies $\mathbf{E}^{(i)}$, see below.
    \item $\mathbf{B}^{(i)}_3$ maps the probability $p( tp^{(i)}_t |tp^{(i)}_{t-1},\pi_t )$ to jump from one topic $tp^{(i)} \in S^{(i)}_3$to another. We consider this transition to be noiseless. This action is only relevant in Simulation 4, in which we compare the model with or without prior information about the topic. Please note that we only use this factor in Simulation 4.
\end{itemize}

The tensor $\mathbf{C}^{(i)}=\mathbf{C}_1^{(i)} \otimes \mathbf{C}_2^{(i)}\otimes \mathbf{C}_3^{(i)}$  encodes the priors over the observations that encode preferred outcomes:
\begin{itemize}
    \item $\mathbf{C}_1^{(i)}$, prior preferences on the outcomes $O_1^{(i)}$, (e.g. for Level 3 encoding sentences, prior on expected words), which is flat in our simulations;
    \item $\mathbf{C}_2^{(i)}$, prior preferences on the outcomes $O_2^{(i)}$ (e.g. for Level 3 encoding sentences, prior on expected word locations), which is flat in our simulations;
    \item $\mathbf{C}_3^{(i)}$, prior preferences on the feedback response $O_3^{(i)}$ (e.g. for Level 3 encoding sentences, a higher prior for having classified the sentence according to the "correct" topic than the "wrong" topic. Please note that we only use this factor in Simulation 4.
    
\end{itemize}

The tensor $\mathbf{D}^{(i)}=\mathbf{D}_1^{(i)}\otimes \mathbf{D}_2^{(i)} \otimes \mathbf{D}_3^{(i)}$ encodes the priors over the hidden states:
\begin{itemize}
    \item $\mathbf{D}_1^{(i)}$, priors over the \emph{Content} at the corresponding level. It corresponds to the initial distribution of the content variable (e.g. at Level 3, it is the prior over the initial sentence distribution). For simplicity, here we use flat priors. However, it is possible to use this feature to model the fact that syllabes, words and sentences have different frequencies.
    \item $\mathbf{D}_2^{(i)}$, priors on the \emph{Locations}. For simplicity, here we assign a very high prior to the first element of the content, following the assumption that people start reading (for example) a word from the first syllable. During reading, after each loop the $\mathbf{D}_1^{(3)}$ is initialized to the next word not yet read.
    \item $\mathbf{D}_3^{(i)}$, priors on the \emph{Topics}. We only use this feature in Simulation 4, in which we show that having prior information about the topic of the sentence can speed up its reading. In all the other simulations, we set flat priors over Topics.  

    The matrix $\mathbf{E}^{(i)}$ encodes the priors over the policies. 

\end{itemize}

\subsection{Saccade selection}

Saccade selection arises from a competition among policies spanning all hierarchical levels, determining the location of the next word in the sentence (Level 3), syllable in the word (Level 2), and letter in the syllable (Level 1). The number of policies at each level varies depending on the simulation. For example, when reading 8-word sentences in Simulation 2, the model uses 8 policies to jump among words, 4 policies to jump among syllables and 5 policies to jump among letters.

Policy selection considers the Expected Free Energy (EFE) described in \eqref{eq:EFE}. In our simulations, we employ flat prior preferences on Content and Location (denoted as $\mathbf{C}^{(i)}_1$ and $\mathbf{C}^{(i)}_2)$, respectively), resulting in policy selection that solely weighs the information gain associated with the following word, syllable, or letter. However, in Simulation 4, policy selection also depends on the greater preference $(\mathbf{C}^{(i)}_3)$ for receiving a "correct" than a "wrong" feedback response (we remind that this factor is absent in the other simulations). 

\begin{figure}
    \centering
    \includegraphics[width=0.75\linewidth]{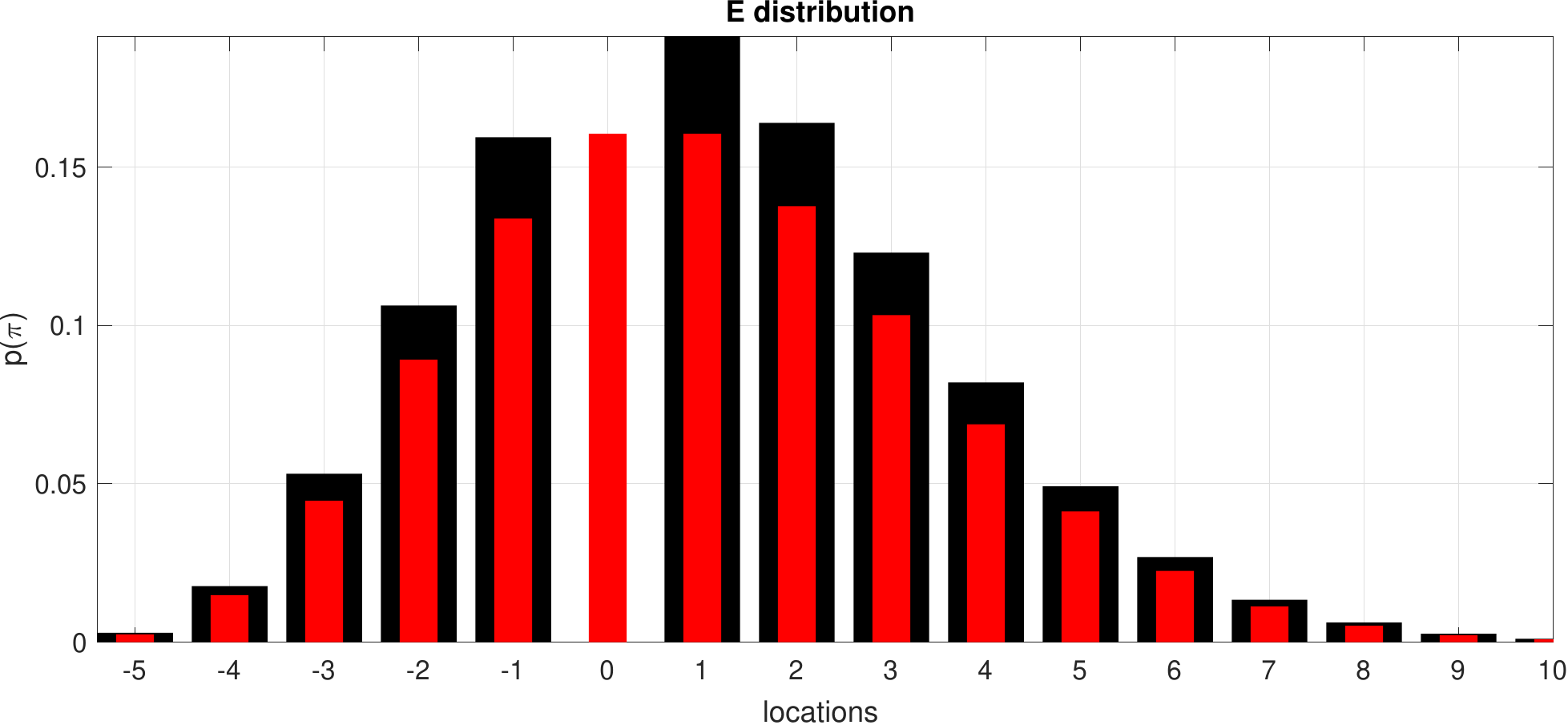}
    \caption{Priors on the policies that make transitions between word locations at Level 3 of the generative model. We first set the prior as a Poisson distribution with $\lambda=6$ (red bars), centered on the current location (i.e., the value of the maximum of the distribution is in 0). We next set the probability of reading the same word two consecutive times to zero and re-normalize the distribution, leading the distribution shown by the black bars.}
    \label{fig:Epriors}
\end{figure}

Furthermore, policy selection considers the prior distribution over policies (denoted as $\mathbf{E}^{(i)}$). Our simulations adopt flat priors for Level 1 ($\mathbf{E}^{(1)}$) and Level 2 ($\mathbf{E}^{(2)}$), signifying no inherent constraints on transitioning between letters within a syllable or between syllables within a word. To discourage overly long saccades, a finite moving window for favored transitions among word locations is set for Level 3 ($\mathbf{E}^{(3)}$). This is achieved through a Poisson distribution $Pois(\lambda)$, where $\lambda$ equals 6 (as displayed in Figure~\ref{fig:Epriors}, depicted by red bars). Additionally, the probability of reading the same word after having already recognized it is set to zero, resulting in the normalization of the distribution (note however that it the distribution allows backtracking to a previous word during reading). This culminates in the distribution depicted by the black bars in Figure~\ref{fig:Epriors}. The chosen distribution enables the creation of a finite window to determine probabilities related to the content under consideration. Iterating this window at each step enables the processing of sentences of arbitrary length.

\begin{figure}
\centering
\includegraphics[width=1\linewidth]{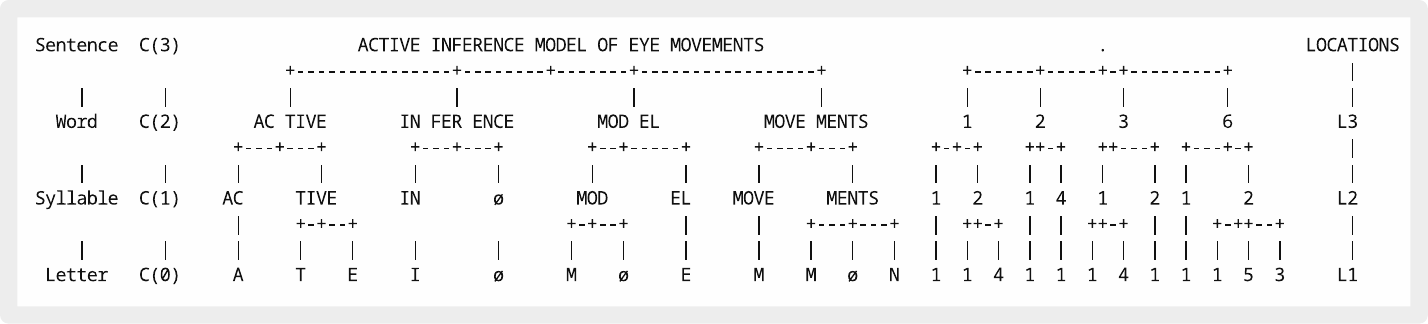}
\caption{Graphical explanation of the model's reading process for the example sentence "Active inference model of eye movements" introduced in Figure \ref{fig:HAI_model}. The figure illustrates the distinction between two factors of the generative model: content (left tree) and location (right tree). Refer to the main text for a detailed explanation.}
\label{fig:TREE}
\end{figure}

Furthermore, Figure~\ref{fig:TREE} presents a schematic illustration of the textual elements inferred by the model while reading the sentence "Active inference model of eye movements" as depicted in Figure~\ref{fig:HAI_model}. The diagram distinguishes between two factors of the generative model -- Content and Location -- illustrated by the left and right trees, respectively. The left (content) tree in Figure~\ref{fig:TREE} demonstrates that the model infers the first word "active" at Level 2 by examining the letter "a". This, in turn, leads to the inference of the syllable "ac" at Level 1, followed by the letters "t" and "e", enabling the inference of the syllable "tive" at Level 2. The right (location) tree in Figure~\ref{fig:TREE} shows the respective locations of these three letters "a", "t", and "e" within their syllables at Level 1, and the positions of "ac" and "tive" within the word at Level 2. This process continues for the subsequent words (excluding "of" and "eye", which are skipped during reading, as seen in Figure \ref{fig:TREE}), until the entire sentence is recognized.

\section{Results}

\subsection{Simulation 1: Reading single words}
\begin{figure}
\centering
\subfloat[]{\includegraphics[width=0.455\textwidth]{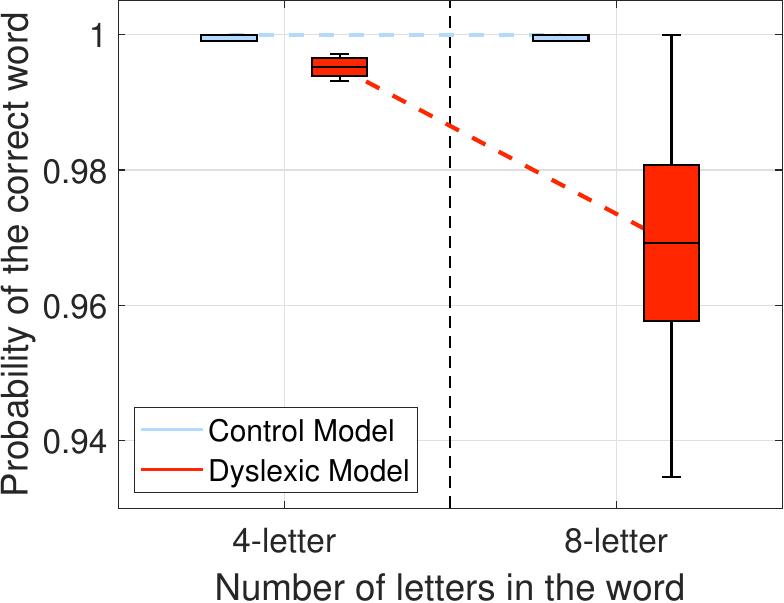}}~~
\subfloat[]{\includegraphics[width=0.455\textwidth]{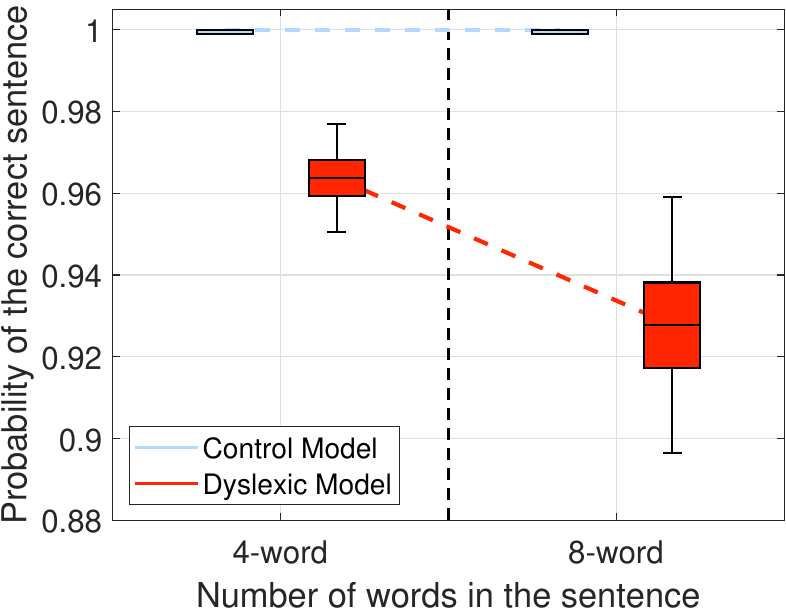}}
  \caption{Results of Simulations 1 and 2: Probabilities assigned to the correct words. (a) Simulation 1: Probabilities assigned by the \emph{Control model (CM)} and the \emph{Dyslexic model (DM)} while reading 4-letter and 8-letter words. (b) Simulation 2: Probabilities assigned by the two models while reading 4-word and 8-word sentences. In this and the subsequent boxplots, the horizontal line represents the mean $\mu$. The edges of the rectangle represent $\mu \pm \sigma$, where $\sigma$ is the standard deviation of the mean. The area within the boxplot above the mean represents the interval between $\mu + \sigma$, while that below the red line represents the interval between $\mu - \sigma$. The vertical black lines limits extend within the range of $\mu \pm 3\sigma$.}
  \label{fig:Zoccolotti_letter_word}
\end{figure}

In this task, the objective is to read and recognize 100 words consisting of either 4 or 8 letters, selected from a pool of 726 words ranging from 1 to 8 letters (see Tab.~\ref{tab:Dictionary_simulation1} for the list of words used in this simulation, all of which are part of the dictionary of BERT). Syllables for each word are generated from words using the methods described in \citep{liang1983word}. For ease of evaluation, each word is assigned the same a-priori probability of 1/726 in the model.

The model's performance is assessed using four metrics: word recognition accuracy, the probability assigned to the correct word, the number of saccades (forward, backward, or total), and their amplitude (i.e., the number of locations between letters). It is important to note that for this specific simulation, only levels 1 and 2 of the model are utilized, while Level 3 is not included.

To better evaluate the model, we compare two versions: (i) a \emph{Control model (CM)} with no noise in its transition functions, allowing it to adequately consider the prior context during reading, and (ii) a \emph{Dyslexic model (DM)}, aiming to capture qualitatively key aspects of this disorder. Previous research has demonstrated that individuals with dyslexia exhibit a reading style characterized by more fragmented and laborious patterns of eye movement, with a higher number of shorter saccades when reading words, and these difficulties are more pronounced with longer words  \citep{zoccolotti2005word,mackeben2004eye,franzen2021individuals}. \textcolor{rev}{To provide a proof of concept that our DM model can replicate this pattern of results, here we follow the proposal that dyslexics struggle to integrate prior linguistic context  \citep{jaffe2015computational} -- and introduce noise in the transition functions of both the first (syllable) level and the second (word) levels. For this, we set $\delta^{(1)}=\delta^{(2)}=0.15$ in the DM model, see Methods section.}


The \emph{CM} demonstrates perfect accuracy (100\%), whereas the \emph{DM} exhibits slightly lower accuracy for both 4-letter (99\%) and 8-letter words (97\%). Additionally, the \emph{CM} assigns a 100\% probability to the correct words, whereas the \emph{DM} assigns a significantly lower probability to the correct word, especially for longer words. Please refer to Figure~\ref{fig:Zoccolotti_letter_word}(a) and Table~\ref{tab:Letter_AccProb} for detailed results.

\begin{figure}
\centering
    \subfloat[]{\label{fig:LetterTotalSacc}\includegraphics[width=0.45\linewidth]{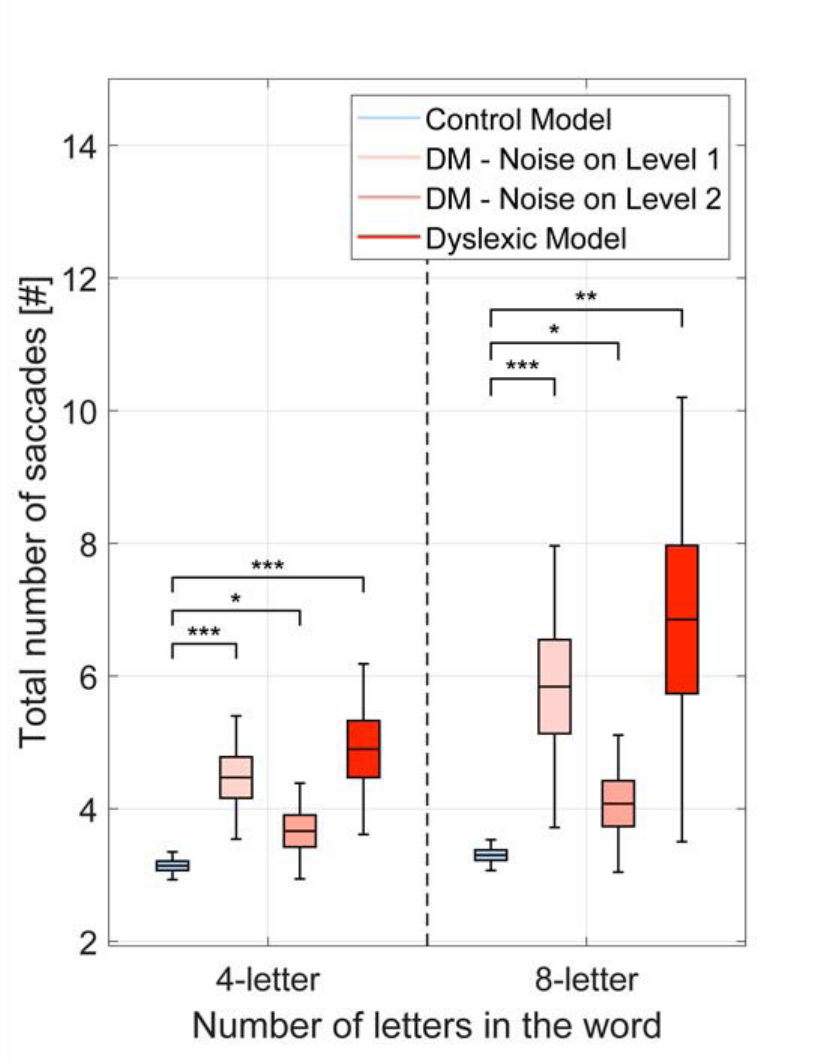}}~~
    \subfloat[]{\label{fig:LetterBackw}\includegraphics[width=0.45\linewidth]{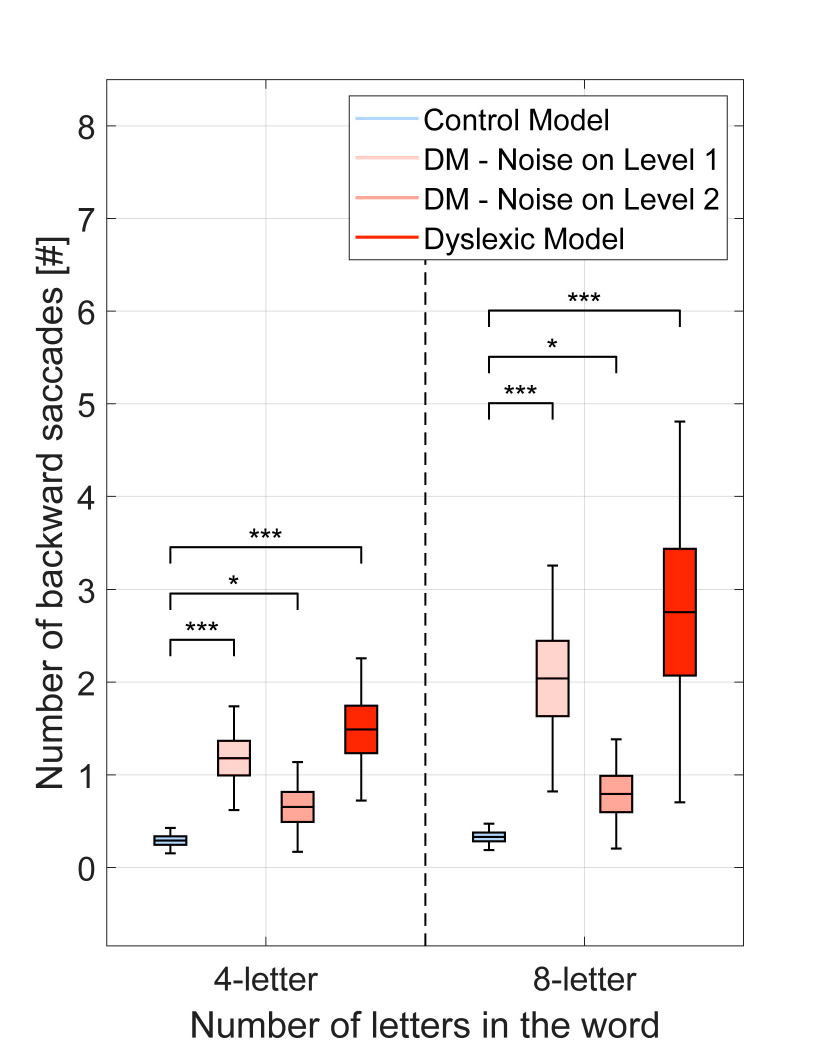}}
    \caption{Simulation 1 results: Total number of saccades (a) and number of backward saccades (b) while reading 4-letter and 8-letter words. The figure compares the \emph{Control Model (CM)} with no noise, the \emph{Dyslexic Model (DM)} with noise at all hierarchical levels, and two versions of the DM, with noise only at the level of syllables (\emph{DM - Noise on Level 1}) or words (\emph{DM - Noise on Level 2}). See Table~\ref{tab:ANOVA_Letter_sacc} and Table~\ref{tab:ANOVA_Letter_backsacc} for detailed results and statistical comparisons.Horizontal bars indicate statistically significant differences from the Control Model (CM). Significance levels are indicated as follows: *~$p < 0.05$, **~$p < 0.01$, ***~$p < 0.001$.}
    \label{fig:Lett}
\end{figure}

Moreover, the \emph{DM} exhibits a significantly higher number of total saccades (Figure~\ref{fig:Lett}(a)) and backward saccades (i.e., saccades to any of the preceding letters, Figure~\ref{fig:Lett}(b)) while reading both 4- and 8-letter words compared to the \emph{CM}. Detailed statistical comparisons are available in Tab. \ref{tab:ANOVA_Letter_sacc} and \ref{tab:ANOVA_Letter_backsacc}. To delve deeper into the impact of noise on eye movement dynamics during word reading, we created two variants of the \emph{DM}, one with noise in the transition functions at the syllable level (\emph{DM} - Noise on Level 1) and another with noise at the word level (\emph{DM} - Noise on Level 2). This analysis reveals that noise at the first (syllable) level has a more significant effect than noise at the second (word) level.

\begin{figure}
  \subfloat[]{\includegraphics[width=0.45\textwidth]{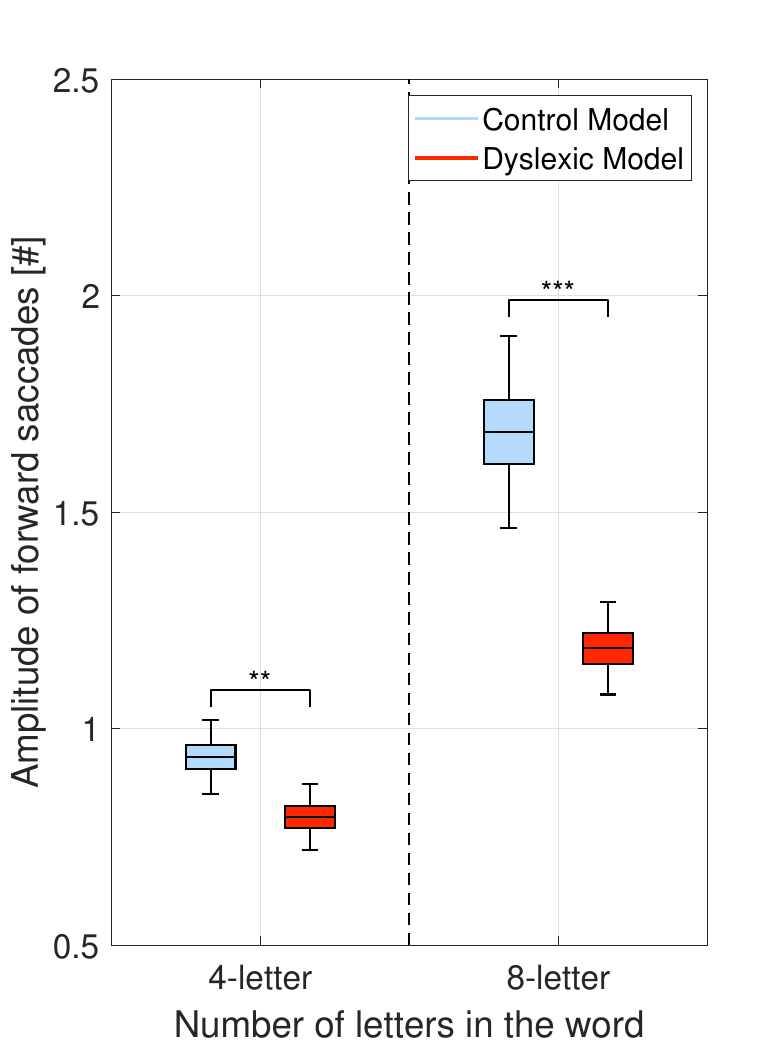}}~~
  \subfloat[]{\includegraphics[width=0.45\textwidth]{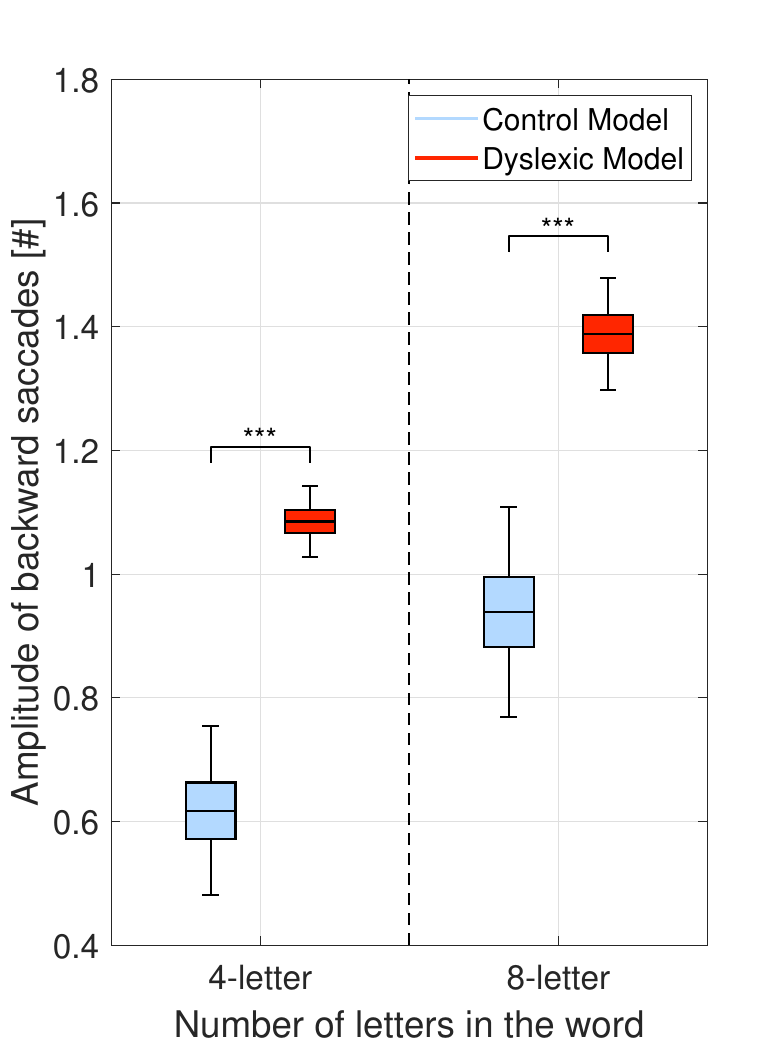}}
  \caption{Simulation 1 results: Amplitude of forward (a) and backward (b) saccades while reading 4-letter and 8-letter words. The figure compares two models: the \emph{Control Model (CM)} with no noise, and the \emph{Dyslexic Model (DM)} with noise at both hierarchical levels. See Table~\ref{tab:ANOVA_Letter_Forw_AMPLITUDE} and Table~\ref{tab:ANOVA_Letter_Backw_AMPLITUDE} for detailed results and statistical comparisons.}
\label{fig:Zoccolotti_letter_amplitude}
\end{figure}

In addition, the \emph{DM} exhibits significantly shorter forward saccades (Figure~\ref{fig:Zoccolotti_letter_amplitude}(a)) but longer backward saccades (Figure~\ref{fig:Zoccolotti_letter_amplitude}(b)) while reading both 4- and 8-letter words in comparison to the \emph{CM}. Detailed statistical comparisons are available in Table \ref{tab:ANOVA_Letter_Forw_AMPLITUDE} and in Table \ref{tab:ANOVA_Letter_Backw_AMPLITUDE}.

In summary, our model successfully reproduces a wide range of empirical findings regarding how dyslexic individuals read single words, encompassing accuracy, reaction times, and eye movements. The lower probability assigned to words by the DM compared to the CM, as depicted in Figure~\ref{fig:Zoccolotti_letter_word}(a), closely mirrors the results reported by \citep[Fig.~1(B)]{zoccolotti2005word}. The accuracy results (Table~\ref{tab:Letter_AccProb}) are also consistent with the finding that dyslexic individuals fail to correctly read a small percentage of single words. Moreover, our simulations reveal that dyslexic individuals read significantly slower than controls  \citep[Fig.~1(A)]{zoccolotti2005word,o2005effect}. This slower reading arises from the DM making more saccades than the CM, as illustrated in Figure~\ref{fig:Lett}, with reading time being proportional to the number of saccades.

Furthermore, the results displayed in Figure~\ref{fig:Zoccolotti_letter_amplitude}(a) closely match the finding that dyslexic individuals make shorter forward saccades compared to controls, particularly evident for longer words  \cite[Fig.~4]{mackeben2004eye}. Lastly, the results presented in Figure~\ref{fig:Zoccolotti_letter_amplitude}(b) closely resemble the finding that dyslexic individuals make significantly more and larger backward saccades than controls during reading  \cite[Fig.~5]{prado2007eye}. Collectively, these outcomes demonstrate that our model effectively captures possible aspects of reading skill differences and/or development, providing valuable insights into the underlying mechanisms of reading disorders like dyslexia.

Significantly, our model offers a mechanistic understanding for all these observed findings, including the fragmented reading style of dyslexics. The DM's shorter forward saccades stem from its poor contextual memory, making it less capable of predicting the next word efficiently. Consequently, it requires more saccades and time to read a text. Additionally, the increased need for backward saccades arises because the DM occasionally has to backtrack in the text to retrieve lost context. Moreover, our model readily explains why dyslexic reading impairments are more pronounced for longer words: noise in the transition function(s) accumulates over time, making reading longer words progressively more challenging. This comprehensive mechanistic explanation underscores the potential usefulness of our model in shedding light on the underlying cognitive processes contributing to dyslexia and its impact on reading behavior.

\subsection{Simulation 2: Reading Sentences}

This task involves reading and recognizing 100 sentences composed of 4 or 8 words (all of which are included in the dictionary of BERT \citep{Devlin20194171}), with each word consisting of 1 to 4 syllables. The sentences with 4 words have an average length of $25.96 \pm 1.96$ letters within the range of $[21,31]$. Similarly, sentences with 8 words have an average length of $50.24 \pm 1.64$ letters, in the interval of $[47,55]$ (see Table \ref{tab:Dictionary_characterLenght}). To make the text challenging, we carefully designed sentences with a substantial word overlap (Table \ref{tab:Dictionary_simulation2}). For this task, we utilize the comprehensive model depicted in Figure \ref{fig:HAI_model}, encompassing three hierarchical levels. At the beginning of the simulation, we use BERT to generate prior probability distributions over upcoming sentences, comprising the next 4 or 8 words. These distributions are continuously updated during reading as new syllables and words are inferred through informative saccades. 

Previous studies have shown that proficient readers can scan lines of text using only a few saccades, while dyslexic individuals exhibit an increased number of shorter saccades  \citep{hutzler2004eye,rayner1983eye,de1999eye} and fewer word skipping occurrences  \citep{bucci2008poor,jainta2011dyslexic,hawelka2010dual} compared to controls. Additionally, their performance decrease and increased number of saccades are influenced by the number of words in the text (see  \citep{de1999eye}, Fig.~2). To verify whether our model accurately replicates these findings, we compare a \emph{Control model (CM)} with no noise and a \emph{Dyslexic model (DM)} with noise introduced in the transition functions of all three hierarchical levels \textcolor{rev}{(setting the parameter $\delta^{(1)}=\delta^{(2)}=\delta^{(3)}=0.15$, see Methods section).}

\begin{figure}
    \subfloat[]{\label{fig:Word_TotalSacc}\includegraphics[width=0.45\linewidth]{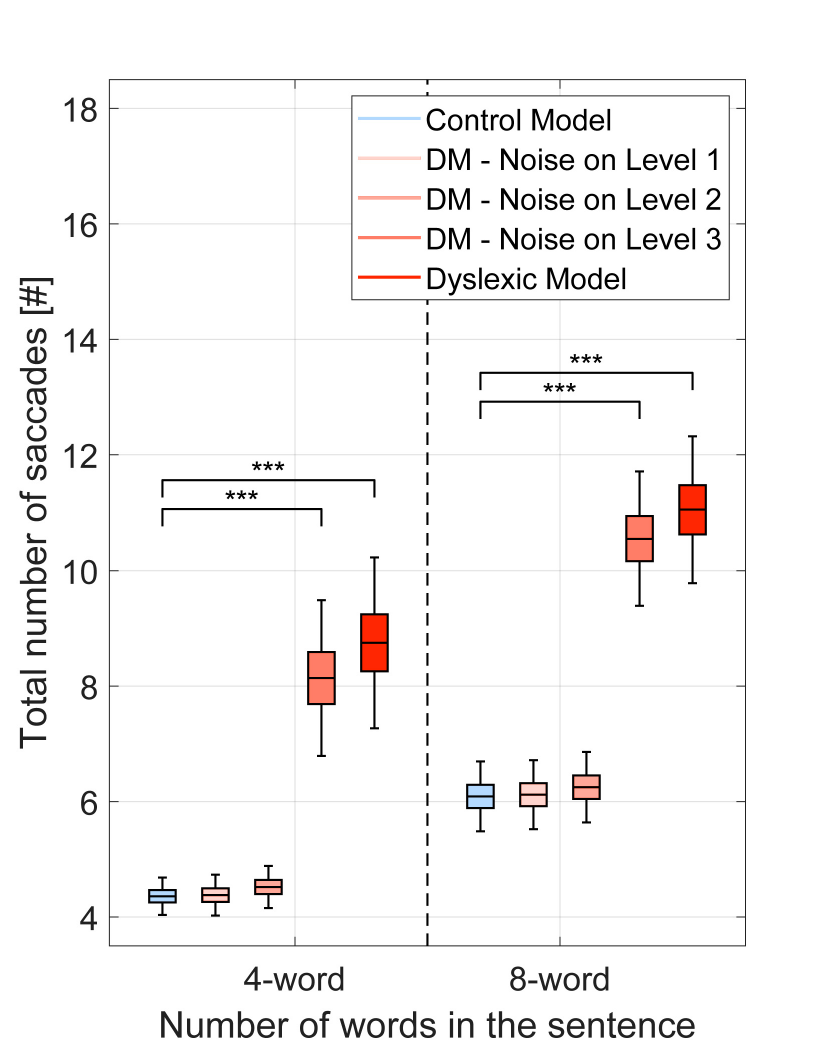}}~~
    \subfloat[]{\label{fig:Word_Back}\includegraphics[width=0.45\linewidth]{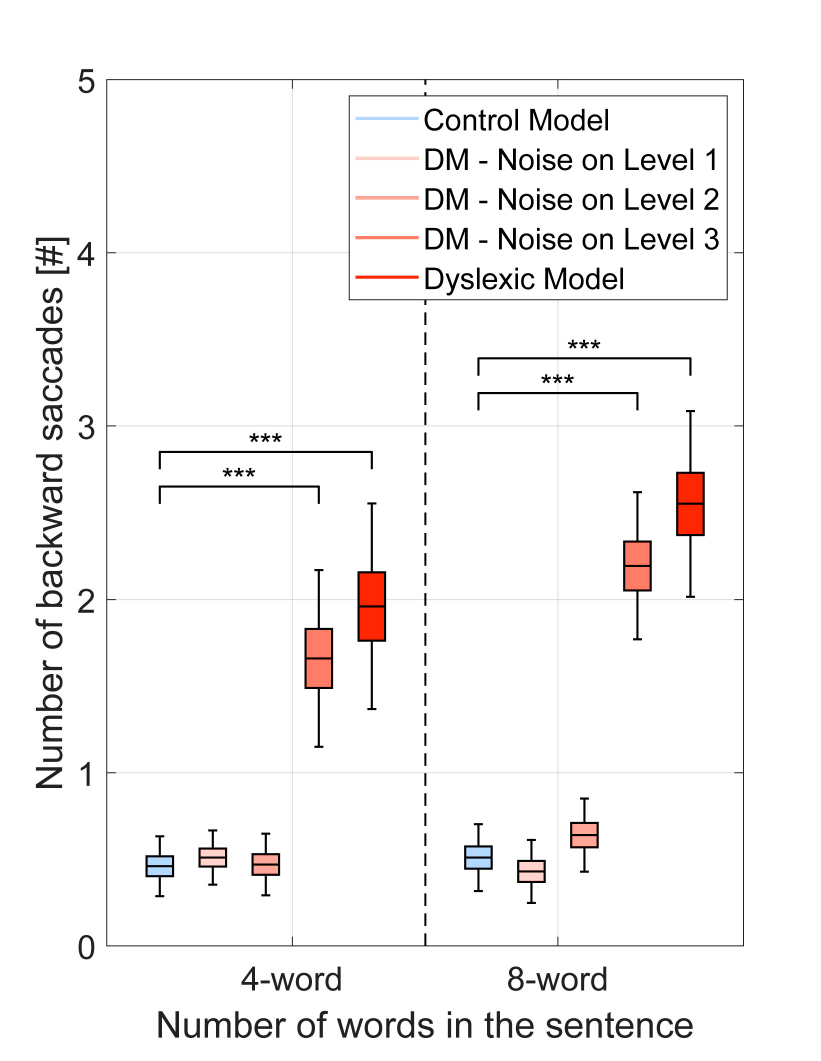}} 
    \caption{Simulation 2 results: Total number of saccades (a) and number of backward saccades (b) while reading 4-word and 8-word sentences. The figure compares the \emph{Control Model (CM)} with no noise, the \emph{Dyslexic Model (DM)} with noise at both hierarchical levels, and three alternative versions of the Dyslexic Model, where we added noise only at the level of syllables (\emph{DM - Noise on Level 1}), words (\emph{DM - Noise on Level 2}), or sentences (\emph{DM - Noise on Level 3}). See Table~\ref{tab:ANOVA_Word_Sacc} and Table~\ref{tab:ANOVA_Word_backsacc} for detailed results and statistical comparisons.}
\label{fig:Word}
\end{figure}

Consistent with the empirical findings, the \emph{DM} assigns significantly lower probabilities to the correct sentences compared to the \emph{CM}, and this effect is more pronounced for longer sentences (Figure~\ref{fig:Zoccolotti_letter_word}(b) and Table~ \ref{tab:Word_AccProb}). Additionally, the \emph{DM} exhibits a significantly higher number of total saccades (Figure~\ref{fig:Word}(a-c)) and backward saccades (Figure~\ref{fig:Word}(b-d)) than the \emph{CM} during sentence reading. When comparing different variants of the \emph{DM} with noise introduced at various levels, we find that noise at the level of sentences has the most significant impact on impairing the reading performance.

\begin{figure}
    \subfloat[]{\includegraphics[width=0.45\linewidth]{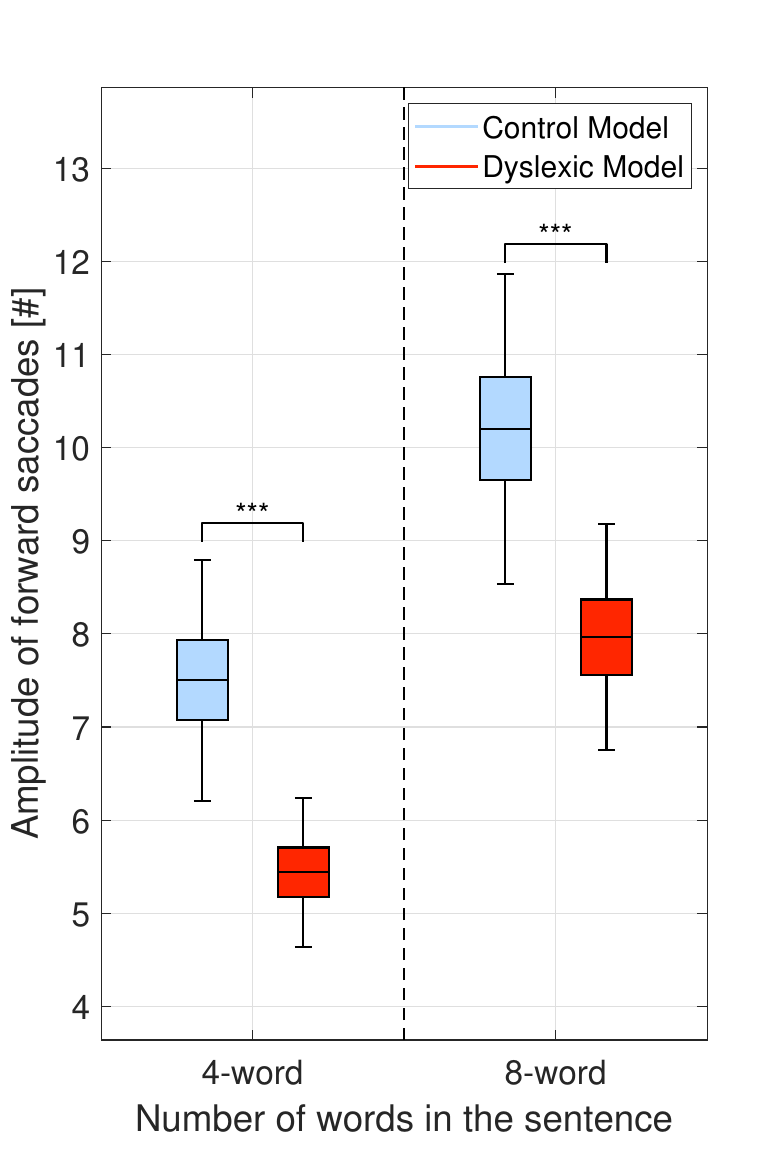}}
    \subfloat[]{\includegraphics[width=0.45\linewidth]{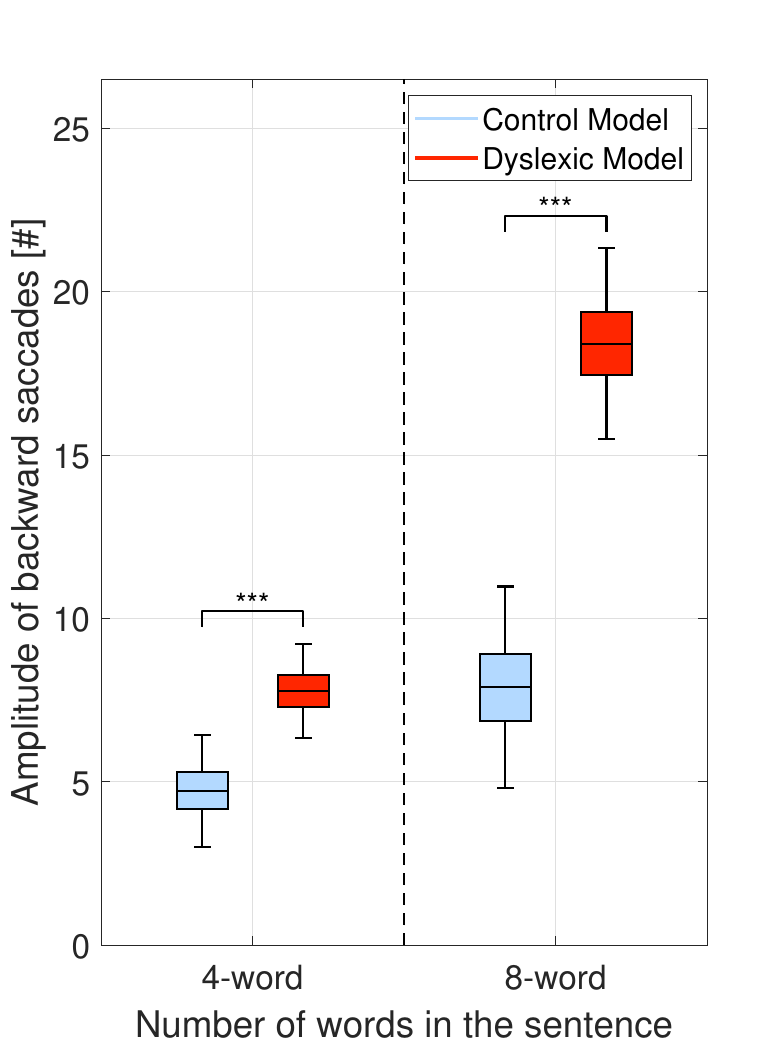}}
    \caption{Results of Simulation 2: amplitude of forward (a) and backward (b) saccades when reading 4-word and 8-word sentences. The figure compares the \emph{Control Model (CM)} with no noise and the \emph{Dyslexic Model (DM)} with noise at all the hierarchical levels. See Table~\ref{tab:ANOVA_Word_ForwAmpl} and Table~\ref{tab:ANOVA_Word_BackwAmpl} for detailed results and statistical comparisons.}
    \label{fig:Zoccolotti_sentence_forward}
\end{figure}

Moreover, the \emph{DM} exhibits significantly shorter forward saccades (Figure~\ref{fig:Zoccolotti_sentence_forward}(a)) and longer backward saccades (Figure~\ref{fig:Zoccolotti_sentence_forward}(b)) when reading both 4-word and 8-word sentences, in comparison to the \emph{CM}. Detailed statistical comparisons can be found in Table~\ref{tab:ANOVA_Word_ForwAmpl} and in Table~\ref{tab:ANOVA_Word_BackwAmpl}.

Collectively, these results build upon those observed for single words (Simulation 1) and successfully replicate empirical findings on dyslexic individuals' reading of sentences. The trends illustrated in Figure~\ref{fig:Word}(a) closely resemble the outcomes reported \citep[Fig.~3(A)]{de1999eye}, which demonstrates that dyslexics make more saccades compared to controls while reading single lines of text (equivalent to reading 8-word sentences in Figure~\ref{fig:Word}(a)). Moreover, the number of saccades exhibited by the DM in our simulations aligns well with the numerical outcomes from the same study. Similarly, the outcomes in Figure~\ref{fig:Zoccolotti_sentence_forward}(a) closely resemble the findings depicted in \citep[Fig.~3(B)]{de1999eye}, revealing that dyslexics make shorter forward saccades than controls. In this context, the amplitudes of both dyslexics and the DM model closely correspond. \textcolor{rev}{However, the larger amplitude of backward saccades observed in our \emph{DM} (Figure \ref{fig:Zoccolotti_sentence_forward}b) is in contrast with the shorter amplitude of regressions observed in dyslexic readers in \citep[Fig. 3d]{de1999eye}.}

\textcolor{rev}{In summary, our results successfully replicate the observation that dyslexic individuals tend to produce a higher number of shorter saccades while reading sentences, and that their reading performance is influenced by the number of words in the sentence (see \cite[Fig. 2]{de1999eye}). The underlying causes of these impairments are similar to those identified in the single-word reading task (Simulation 1), but their effects are amplified in sentence reading, where comprehension depends on integrating information over longer timescales. This also helps explain why noise at the sentence level significantly disrupts task performance (Figure~\ref{fig:Word}). While these qualitative results are promising, further work is needed to achieve a closer—and more quantitative—alignment between the model and empirical findings on dyslexia, as well as to calibrate model parameters appropriately. For instance, our DM generates a relatively high number of regressive (backward) saccades. This occurs because the model often loses contextual information and must backtrack to earlier words in the sentence to reestablish it. However, this pattern is not commonly observed in dyslexic readers (\citep[Fig. 3d]{de1999eye}), and the reasons for this discrepancy remain to be explored.}


\subsection{Simulation 3. Reading novel words and sentences}

Up to this point, our simulations involved reading \emph{known} words and sentences, already present in the BERT vocabulary. Hence, the model could confidently assign probabilities at the second and third layers. Nevertheless, our model can also read \emph{unknown} words and sentences, despite its inability to assign them a prior probability. When confronted with unknown words or sentences, the model reads syllable-by-syllable (or word-by-word) and simultaneously acknowledges uncertainty at the second (or third) level. This mode aligns with the nonlexical (or sublexical) route in dual-route theories of reading  \citep{coltheart2005modeling}.

\color{rev} We conducted a simulation consisting of 100 trials, each involving the reading and recognition of a four-word sentence, designed for comparison with the corresponding tests in Simulation 2. In each trial, one of the four words was unknown—specifically, it was removed from the model’s vocabulary. This resulted in a set of 100 unique unknown words, each ranging from 1 to 4 syllables in length (see Table~\ref{tab:Dictionary_simulation3} for the complete list). The sentences had an average length of $25.46 \pm 1.92$ letters, with a range of $[22, 32]$. A two-sample t-test comparing sentence lengths between this set and the set with only known words used in Simulation 2 revealed no significant difference ($p \approx 0.27$), confirming that the two sets were well matched in terms of both word and sentence length. Trials were terminated upon either successful completion of the sentence or failure of the model to recognize a word. The simulation results indicate that the total number of saccades increased substantially when reading sentences containing unknown words (mean $\approx 12.04$) compared to Simulation 2 (mean $\approx 4.36$); see Table~\ref{tab:unknown} for statistical comparisons. Furthermore, while sentence recognition at Level 3 in Simulation 2 was nearly perfect ($p(C^3|O) \approx 1$), it dropped to approximately $0.49$ (on average) in the presence of unknown words, indicating a failure to correctly recognize the sentence.

\begin{table}[ht!]
\caption{\textcolor{rev}{Simulation 3. Comparison of two conditions: reading 100 sentences from Simulation 2 (4 words – all known) versus 100 sentences in which one word was unknown, i.e., removed from the model’s vocabulary (4 words – 1 unknown). The table summarizes the comparison between these two conditions, reporting the mean, standard error, and p-value of t-tests for three key measures: sentence length, total number of saccades during reading, and the probability of sentence recognition.}}
\vspace{1em}
\scalebox{0.95}{
\begin{tabular}{l|c|c|c}
\cline{2-4}
\multicolumn{1}{l}{} & \textbf{Sentence Length} & \textbf{Total number of Saccades} & \textbf{Probability} \\
\hline
\multicolumn{1}{c|}{4 words – Known}     & $25.96 \pm 1.96$         & $4.36 \pm 0.11$          & $1.00$ (SE $< 10^{-15}$)   \\
\multicolumn{1}{c|}{4 words – 1 Unknown} & $25.46 \pm 1.92$         & $12.04 \pm 0.44$         & $0.49 \pm 0.04$            \\
\hline
\multicolumn{1}{c|}{\textbf{p-value}}    & $p \approx 0.27$         & $p < 0.001$              & $p < 0.001$                \\
\hline
\end{tabular}}
\label{tab:unknown}
\vspace{1em}
\end{table}

\color{black}

\begin{figure}
    \subfloat[]{\includegraphics[width=1\textwidth]{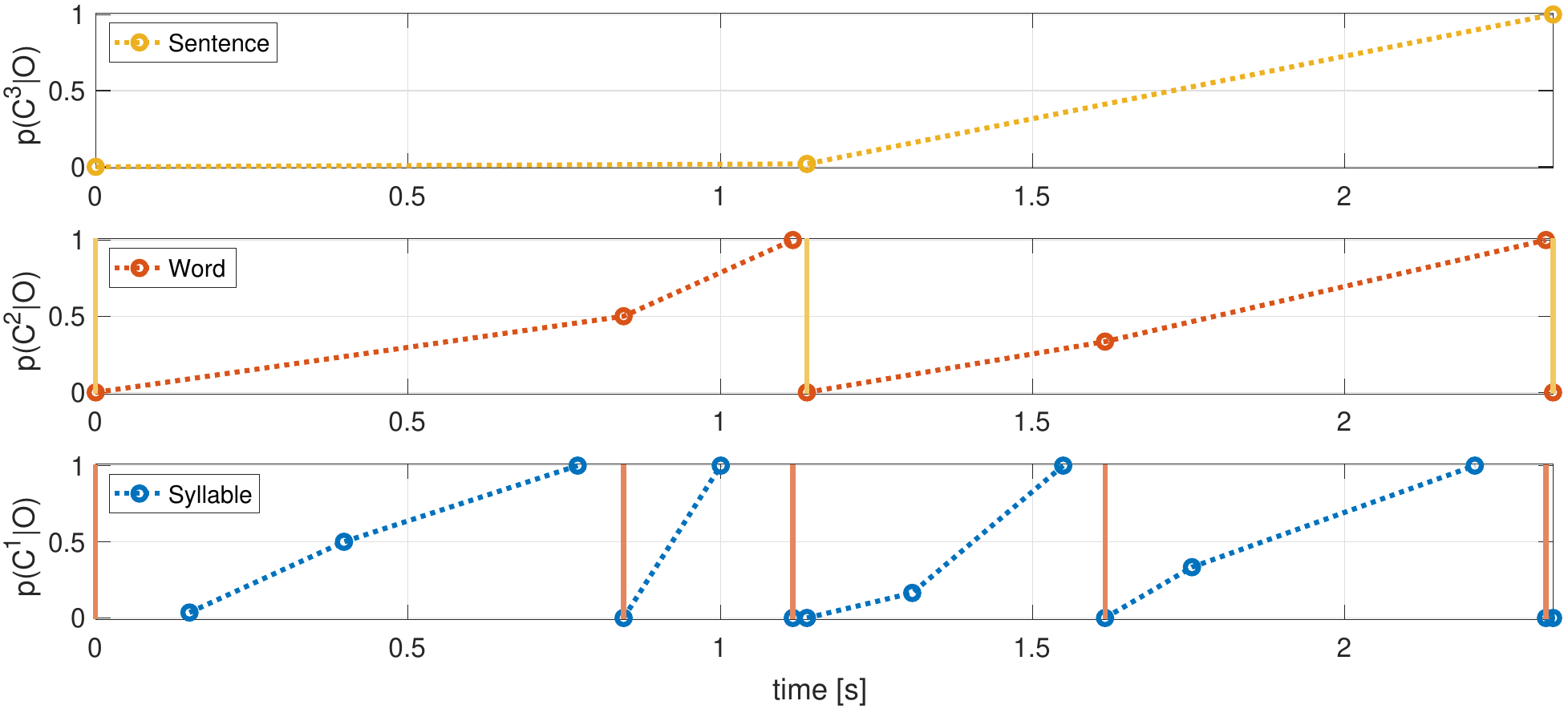}}
    
    \subfloat[]{\includegraphics[width=1\textwidth]{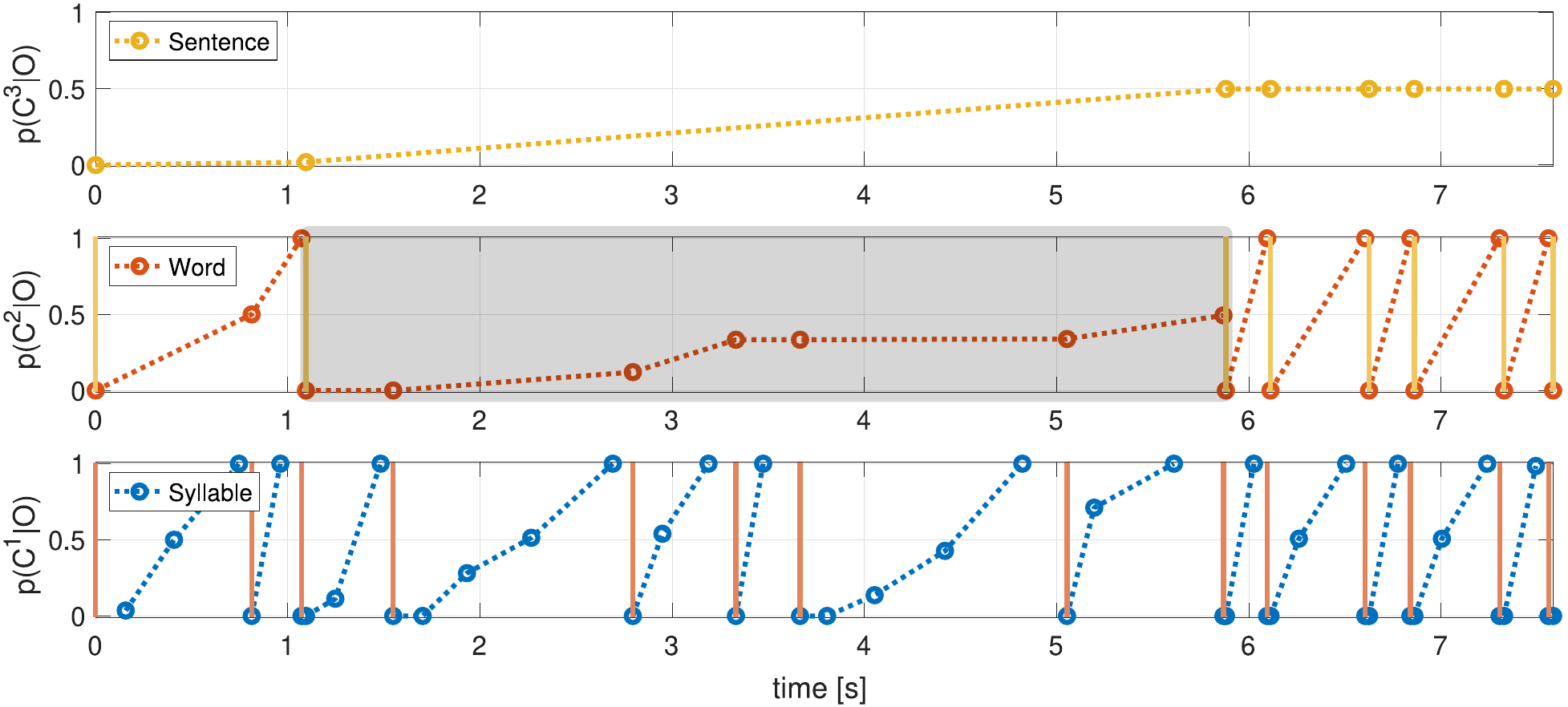}}

    \caption{Simulation 3 results: Reading known vs. unknown words. Evolution of the probabilities of correctly recognized syllables (blue lines), words (orange lines), and sentences (yellow lines) over time, while reading \textcolor{rev}{the final four words of the sentence \emph{This paper is also framed in an offbeat manner}, in two cases: (a) when all the words are known and (b) when the word \emph{offbeat} is unknown (i.e., removed from the vocabulary).} The blue circles at the bottom level indicate saccades. The figure format is consistent with Figure~\ref{fig:HAI_model}(d).}

\label{fig:unknown}
\end{figure}

\textcolor{rev}{To illustrate the model’s behavior, we describe a representative example involving the reading of the final four words of the sentence \emph{This paper is also framed in an offbeat manner}, comparing two versions of the model: one in which} all the words exist in the model's vocabulary (Figure~\ref{fig:unknown}(a)) and another in which the word \emph{offbeat} \textcolor{rev}{was removed} from the vocabulary (Figure~\ref{fig:unknown}(b)). In the version with all known words, the model behaves similarly to the previous simulations, accurately recognizing all the words and the sentence. This is evident in Figure~\ref{fig:unknown}(a), where probabilities at Levels 2 and 3 quickly converge to one after a few iterations.

Conversely, the version without knowledge of the word \emph{offbeat} repeatedly makes saccades to the syllables of this word (at Level 1) until it reaches the maximum number of iterations set for this simulation. It correctly recognizes each syllable and could therefore potentially read the word aloud "non-lexically" (and \textcolor{rev}{potentially} add it to its vocabulary) \citep{coltheart2005modeling}. However, it does not recognize the word (at Level 2) or the sentence (at Level 3). \textcolor{rev}{As shown in Figure~\ref{fig:unknown}(b), both the probability of the unknown word at Level 2 and the probability of the sentence at Level 3 remain below 0.5, indicating a failure of accurate recognition. After failing to recognize the unknown word (\emph{offbeat}), the Level 2 inference settles on the most probable alternative in the model’s vocabulary (e.g., \emph{upbeat}). Unlike the simulation described earlier, here the inference process continues even after an unrecognized word. As a result, the model is able to correctly identify the final word in the sentence (\emph{manner}).}


The example in Fig.~\ref{fig:unknown} illustrates the distinction between reading a known versus a novel word and the termination conditions for inference at each hierarchical level. The inference can terminate in two ways: first, when the model reaches a threshold of confidence about the to-be-recognized syllable, word, or sentence (\textcolor{rev}{i.e., the Expected Ambiguity over hidden states of Equation \ref{eq:EFE} falls below a fixed threshold, $\chi^{(i)}=1/8$)}. This happens when the model successfully infers a known syllable, word, or sentence.

Second, the inference can terminate when it reaches a maximum number of iterations $K_{\max}^{(i)}$ at each $i$-th level. This sets the maximum number of times Level 1 can jump between letters to recognize a syllable, the maximum number of times Level 2 can jump between syllables to recognize a word, or the maximum number of times Level 3 can jump between words to recognize a sentence. In our simulations, for illustrative purposes, we set $K_{\max}^{(i)}$ to very high values ($K_{\max}^{(3)} = 7$ and $K_{\max}^{(2)} = K_{\max}^{(1)} = 6$) but these values can be fine tuned to fit individual participants or participant groups. 

As shown in Figure~\ref{fig:unknown}, Level 1 successfully infers all the syllables within 3 iterations and never reaches $K_{\max}^{(1)}$. However, Level 2 fails to recognize the unknown word \emph{offbeat}: it continues jumping between the syllables \emph{off} and \emph{beat} until it reaches the maximum number of allowed iterations (6 iterations, indicated in the grey panel). When the Level 2 inference of the current word is halted, a low confidence message is reported to the level above \textcolor{rev}{while inference may still proceed with the next words} 
(that are successfully recognized). Similarly, Level 3 fails to infer the unknown sentence after the maximum number of iterations (7 iterations), despite correctly recognizing most of the words. At the end of the simulation, the model identifies that the cause of failure is at Level 2, as Level 1 correctly recognized all the syllables. Consequently, the model reads the unknown word syllable-by-syllable, i.e. using the nonlexical route in dual route theories of reading  \citep{coltheart2005modeling}.


\subsection{Simulation 4: Knowing the topic improves sentence recognition}

In this section, we showcase another capability of the model that has not been discussed yet. The model can be enriched with \emph{priors} regarding the linguistic topic at each hierarchical level, such as the \emph{topic of the sentence} at the third level. This prior knowledge makes certain sentences a-priori more (or less) likely, thus expediting the recognition process if the context aligns correctly.

To illustrate this, we conduct a task involving the recognition of 100 sentences, each containing 9 words, generated by BERT. These sentences comprise an average of $69.55 \pm 5.02$ letters, ranging between 61 and 82 characters (see Table \ref{tab:Dictionary_characterLenght}). We allocate one-third of the sentences to each of the three topics of the European Research Council (ERC): Physical Sciences and Engineering (PE), Life Sciences (LS), and Social Sciences and Humanities (SH); see Table \ref{tab:Dictionary_simulation4} for a full specification of the sentences and their respective topics. Please note that in this simulation, we only use topics at the third level.

\begin{figure}
\centering
    \subfloat[]{\label{fig:Prior_Sacc}\includegraphics[width=0.45\linewidth]{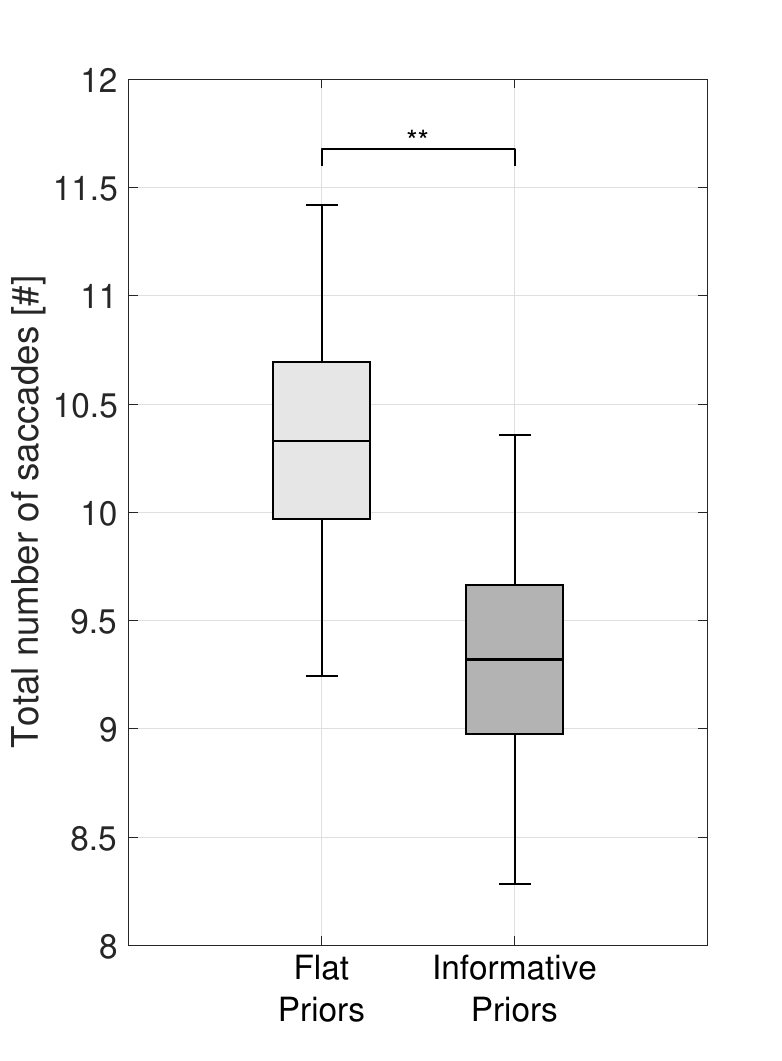}}
    \subfloat[]{\label{fig:Prior_Back}\includegraphics[width=0.45\linewidth]{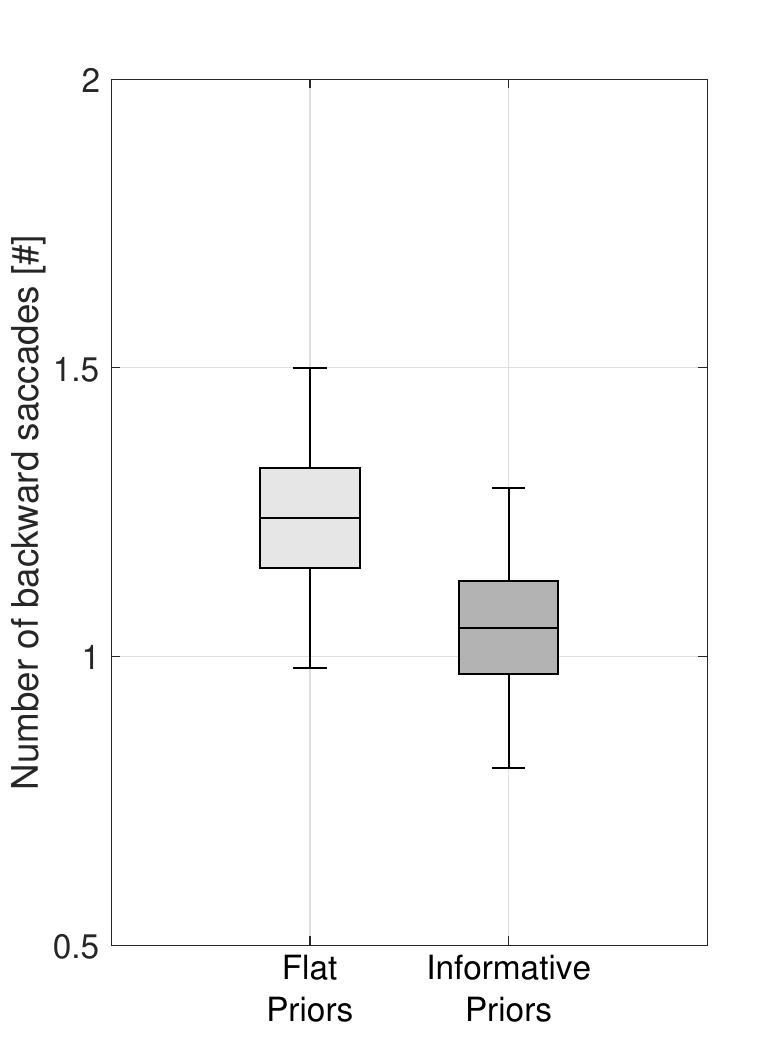}}       
    \caption{Simulation 4 results: How prior information affects the reading process. The figure displays the total saccades (a) and backward saccades (b) of two model versions: one with a flat prior about the topic of the text (flat prior) and another with certain knowledge of the topic (informative prior). The comparison highlights the impact of prior information on the number and nature of saccades during reading. See Table~\ref{tab:ANOVA_Letter_Forw_AMPLITUDE_SIM4} and Table~\ref{tab:ANOVA_Letter_Backw_AMPLITUDE_SIM4} for detailed results and statistical comparisons.}
    \label{fig:Context}
\end{figure}

We compare the performance of two versions of the model: one with a \emph{flat prior} (33.3\%) regarding the sentence being read, and another with an \emph{informative prior} that assigns a probability of 100\% to the correct topic. As depicted in Figure~\ref{fig:Context}, the model with an informative prior requires significantly fewer total saccades and backward saccades than the model with a flat prior (see Table \ref{tab:ANOVA_Letter_Forw_AMPLITUDE_SIM4} and Table \ref{tab:ANOVA_Letter_Backw_AMPLITUDE_SIM4} for statistical comparisons). This advantage arises because knowing the topic allows the model to predict certain words with high confidence, enabling it to skip unnecessary saccades during the reading process. Intriguingly, despite the model with an \emph{informative prior} skipping several words, it does not necessitate more backward (corrective) saccades compared to the model with a \emph{flat prior}. This indicates that the prior information promotes a better speed-accuracy trade-off.

\section{Discussion}

In this paper, we presented a novel hierarchical active inference model of reading, which combines the robust generative abilities of large language models \citep{Devlin20194171,radford2019language} with the inferential and information-gathering capacities of active inference \citep{friston2010free,parr2022active}.

Large language models, like BERT \citep{Devlin20194171} and GPT (Generative Pre-trained Transformer)  \citep{radford2019language}, have demonstrated remarkable performance in various natural language processing tasks, including question answering, language translation, text classification, and sentence prediction. These models can learn essential linguistic structures without external supervision, making them valuable tools to explore language processing in the brain  \citep{manning2020emergent,goldstein2022shared}. However, they lack the \emph{active} strategies employed by human readers, such as the ability to make saccades selectively on the most informative parts of the text  \citep{ferro2010reading,norris2006bayesian}.

To capture active reading, we integrate BERT \citep{Devlin20194171} into a hierarchical active inference framework \citep{friston2010free}. Active inference is a theoretical framework that describes how the brain employs probabilistic inference over a generative model to effectively sample sensory inputs, minimizing variational free energy or prediction errors -- and that has been applied too many tasks, such as perceptual processing, goal-directed navigation, robot control, and social interaction \citep{hwang2018dealing,maisto2023interactive,oliver2021empirical,van2023bridging,priorelli2023modeling,parr2022active}. For modeling reading, we utilize a hierarchical generative model, aligning with evidence suggesting that language processing in the brain operates through hierarchically organized predictions and prediction errors \citep{schmitt2021predicting,caucheteux2023evidence,heilbron2022hierarchy}.

By placing BERT at the top of the hierarchy, we ensure accurate next-word predictions and the model's capability to tackle real-world reading tasks. Beyond next-word prediction, our hierarchical model facilitates both inferring the content being read and predicting future elements across different linguistic levels, such as syllables, words, and sentences. Our simulation results demonstrate the model's perfect accuracy in inferring words (Simulation 1) and sentences (Simulation 2) during reading. Additionally, this hierarchical structure enables the model to read novel words not present in BERT's vocabulary (Simulation 3), assembling them syllable-by-syllable, representing the nonlexical (or sublexical) route in dual route theories of reading \citep{coltheart2005modeling}. Moreover, the model can incorporate prior knowledge, like the topic of the sentence, to expedite the reading process (Simulation 4). Significantly, the model offers interpretability through its alignment of distinct levels with established linguistic constituents (sentences, words, syllables), which could correspond to distributed neural populations or cell assemblies within the brain \citep{heilbron2022hierarchy,pulvermuller1999words}. While delving into the neural mechanisms that underlie this proposed framework extends beyond the confines of this article, future studies could encompass the simulation of neuronal dynamics during the inference process, in terms of variational free energy minimization \citep{friston2017active,isomura2022experimental}.
 
Using active inference, our model characterizes reading as an \emph{active}, hypothesis testing process, allowing us to simulate eye movements during reading. The core idea is that saccades are directed to the most informative parts of the text, enabling the model to test its predictions and reduce uncertainty about the text being read  \citep{ferro2010reading,norris2006bayesian,donnarumma2017action,friston2012perceptions}. Remarkably, this \emph{epistemic} objective, aimed at uncertainty reduction, emerges naturally from active inference, as explained in Section \ref{sec:methods}. In summary, our model offers a comprehensive account of prediction-based written text processing, involving hierarchical inference of the content being read, prediction of upcoming textual elements, and active testing of predictions through saccades, which, in turn, inform further inference.

Interestingly, our model also enables the simulation of abnormal eye movement patterns observed in reading disorders such as dyslexia \citep{maisog2008meta,horowitz2014reading,price2012review}. While the simulations presented here serve as a proof of concept, they qualitatively replicate empirical findings showing that individuals with dyslexia, compared to neurotypical readers, exhibit fragmented text processing. This is characterized by an increased number of shorter saccades during the reading of both single words (Simulation 1) and sentences (Simulation 2) \citep{zoccolotti2005word,de1999eye,hutzler2004eye,rayner1983eye}. Our model attributes these reading deficits to a specific disorder of predictive processing: namely, an attenuation of the influence of prior information during inference \citep{jaffe2015computational}. By introducing noise into the model’s transition functions, we impair its working memory of prior inference steps, thereby forcing it to repeatedly gather evidence. This mechanism, which reflects the cumulative impact of noise on inference, accounts for why perceptual difficulties in dyslexia become increasingly pronounced with longer words and more complex sentence structures \citep{zoccolotti2005word,de1999eye}.

However, it is crucial to note that our simulations only aim to provide a proof of concept that disorders of reading such as dyslexia can be aligned with hierarchical predictive processing theories. Dyslexia is highly heterogeneous, and its underlying causes are still heavily debated  \citep{perry2019understanding,zorzi2012extra,ramus2003developmental,stein1997see,tallal2004improving,ziegler2020learning}. It is unlikely that the attenuation of priors  \citep{jaffe2015computational} explored in this article would provide a comprehensive account of all aspects of dyslexia. However, our model is easily extendable to incorporate other (non-alternative) mechanisms that could contribute to reading impairments. For instance, various proposals point to disorders of low-level information sampling and attention shifting \citep{hari2001left}, or a visual attention span disorder characterized by a reduction in the number of letters that can be processed in parallel \citep{prado2007eye}. Additionally, errors in eye movements (overshooting and undershooting) can be incorporated, by introducing noise in the (transition) function mapping eye movements to subsequent spatial positions. These proposals (and others) can be readily integrated into the model by modifying the likelihood function that maps letters to syllables, as described in Section~\ref{sec:methods}. \textcolor{rev}{Furthermore, while this study used predefined levels of noise in the transition models, these parameters can be fit to human data. Such extensions may pave the way for more comprehensive investigations of dyslexia within the predictive processing framework.} 

A limitation of this study is that for simplicity, it uses a relatively simple LLM: BERT \citep{Devlin20194171}. We selected BERT because it is computationally efficient and provides easy access to explicit probability distributions over the next words and sentences. Future work could adopt more advanced LLM models, to provide more robust linguistic predictions. Furthermore, we made some simplifying assumptions. For example, we assumed that syllables, words and sentences are represented as sequence of letters and that reading uses an absolute letter position coding. Future work could investigate alternative assumptions, see for example \citep{snell2022relative} for a discussion of the absolute versus relative position coding in reading. Furthermore, future work could replace the the priors over policies that make transitions between word locations illustrated in Figure \ref{fig:Epriors} with a more realistic distribution derived from empirical data. Future studies could also explore more systematically the role of the different model parameters and hyperparameters (e.g., word and sentence length, amount of noise) and address more complex linguistic datasets.

Additionally, future work could compare side by side the model presented here with established models in the field, such as connectionist personalized models, which have been very successful in explaining deficits observed in dyslexia \citep{perry2019understanding,ziegler2020learning}, and models that simulate eye movements, such as E-Z Reader \citep{reichle1998toward}, SWIFT \citep{engbert2005swift,seelig2020bayesian,rabe2021bayesian}, Über-Reader \citep{reichle2021computational,veldre2020towards}, Glenmore \citep{reilly2002glenmore,reilly2006some}, SEAM \citep{rabe2024seam}, the rational model of eye movements \citep{bicknell2010rational}, and OB1-Reader \citep{snell2018ob1}. By fitting specific parameters of the presented model (e.g., different levels of noise in the transition and/or likelihood functions), it would be possible to develop personalized models (or computational phenotypes \citep{schwartenbeck2016computational}), potentially shedding light on the mechanisms underlying reading difficulties and informing the development of effective interventions to improve reading skills.

The model presented here could be expanded to capture hierarchically higher and more abstract aspects of language processing, such as those related to shared narratives, frames and scripts, which characterize our cultural niches \citep{minsky1986society,schank2013scripts}. When opportunely placed within hierarchical architectures, these higher-level constructs could produce a cascade of predictions that contextualize and guide lower-level linguistic predictions and eye movements \citep{bouizegarene2024narrative,friston2024designing,pezzulo2025predictive}. The presence of these additional constructs creates a complex interplay between prior predictions at different hierarchical levels -- for example, those induced by scripts, statistical regularities of sentences and word frequencies -- and between prediction errors arising at different levels, which remain to be investigated in future studies.

\textcolor{rev}{In principle, the model presented here could also be extended to investigate developmental aspects of language and reading acquisition. Using a pretrained LLM such as BERT can be interpreted as mimicking the capabilities of adult readers. To simulate developmental trajectories, one could impose constraints on BERT's vocabulary or its ability to generate accurate sentence completions. Limiting the vocabulary would force the model to rely more heavily on syllable-by-syllable reading, engaging a non-lexical processing route. Reducing prediction accuracy would result in less confident priors, leading the model to skip fewer words, commit more reading errors, and engage in more fixations and backward saccades. Whether a model with these constraints can successfully replicate the patterns observed during reading development remains to be evaluated in future studies.}

\textcolor{rev}{Another important direction for future research is the integration of more realistic perceptual modules. This would allow the model to capture the complexity of the feature extraction process (e.g., \citep{testolin2017letter}), the parallel processing of multiple letters (e.g., \citep{hannagan2021emergence}), and other perceptual factors known to influence dyslexia, such as the beneficial effects of increased letter spacing (e.g., \citep{zorzi2012extra}). A possible approach to realize this integration is to employ neural networks as front-end perceptual modules that provide probabilistic outputs (e.g., categorical distributions over letters or syllables), which can then be incorporated as likelihoods within a Bayesian model \citep{george2017generative,sancaktar2020end,priorelli2023flexible}. }

Finally, it is worth noting that the hierarchical control architecture proposed here can be extended to a wide range of cognitive tasks beyond reading. Many such tasks can be naturally decomposed in hierarchical control terms by considering the separation of timescales between slower, higher-level perceptual and action processes and faster, lower-level ones \citep{botvinick2009hierarchically,pezzulo2018hierarchical,koechlin2007information,friston2008hierarchical}. For instance, spatial navigation often involves high-level planning over goals or subgoals and low-level motor actions required to reach them \citep{van2023bridging,donnarumma2016problem,tomov2020discovery,donnarumma2017action}. Similarly, action observation and understanding can be structured hierarchically by distinguishing between distal goals, proximal intentions, and immediate motor acts \citep{proietti2023active}. Future work could adapt the hierarchical framework introduced here to explore these and other cognitive domains, shedding light on how flexible, multi-timescale processing supports complex behavior.







\section*{Acknowledgments} This research received funding from the European Research Council under the Grant Agreement No. 820213 (ThinkAhead), the Italian National Recovery and Resilience Plan (NRRP), M4C2, funded by the European Union – NextGenerationEU (Project IR0000011, CUP B51E22000150006, “EBRAINS-Italy”; Project PE0000013, “FAIR”; Project PE0000006, “MNESYS”), and the Ministry of University and Research, PRIN PNRR P20224FESY, PRIN 20229Z7M8N and PRIN 2020529PCP. The GEFORCE Quadro RTX6000 and Titan GPU cards used for this research were donated by the NVIDIA Corporation. We used a Generative AI model to correct typographical errors and edit language for clarity.

\bibliography{HAI_Reading}
\bibliographystyle{plain}

\newpage
\appendix
\pagestyle{plain}

\setcounter{page}{1}

\setcounter{figure}{0} 
\setcounter{table}{0} 

\renewcommand\thefigure{S.\arabic{figure}}      
\renewcommand\thetable{S.\arabic{table}}

\section*{\centering \huge{Supplementary materials}}
\vspace{10em}
\section*{\centering \LARGE{Integrating Large Language Models \\ and Active Inference \\to understand eye movements \\ in Reading and Dyslexia}}

\vspace{4em}

\section*{
\begin{center}
\href{https://orcid.org/0000-0003-4248-5360}{Francesco Donnarumma} \\
\href{https://orcid.org/0000-0003-4565-2769}{Mirco Frosolone} \\
\href{https://orcid.org/0000-0001-6813-8282}{Giovanni Pezzulo}
\end{center}}

\vspace{4em}
\section*{
\begin{center} 
\small {Institute of Cognitive Sciences and Technologies\\ National Research Council \\ Via Gian Domenico Romagnosi, 18A \\00196 Rome Italy}
\end{center}}
\vspace{4em}

\section*{
\begin{center}
\small {Corresponding author: Giovanni Pezzulo E-mail: \href{mailto:giovanni.pezzulo@istc.cnr.it}{\texttt{giovanni.pezzulo@istc.cnr.it}}}
\end{center}}
\pagebreak

\subsection*{Supplementary info}
In this document, we provide the tables of statistics from the simulations presented in the main article (found in Supplementary Tables section) and the Datasets of sentences and words used in the simulation (found in Datasets section).

\subsection*{Supplementary tables}

Below are the supplementary tables presenting statistics related to the simulations discussed in the main manuscript.

\subsection*{Supplementary tables for Simulation 1}

Table \ref{tab:Letter_AccProb} displays the accuracy and probability assigned to the correct words.  Table \ref{tab:ANOVA_Letter_sacc} shows the T-tests on total saccades and Table \ref{tab:ANOVA_Letter_backsacc} the backward saccades. Table \ref{tab:ANOVA_Letter_Forw_AMPLITUDE} and Table \ref{tab:ANOVA_Letter_Backw_AMPLITUDE} presents the T-tests for the amplitude of forward and backward saccades respectively. The $p$-values lower to $0.05$ are indicated in bold.
\begin{table}[ht!]
\centering
\caption{Simulation 1: Reading words of 4 or 8 letters. The table presents the accuracy and the probability assigned to the correct words by 4 different models: the \emph{Control model} (CM), the \emph{Dyslexic model} (DM) variant with noise at Level 1 only, the \emph{Dyslexic model} (DM) variant with noise at Level 2 only, and the \emph{Dyslexic model} (DM) with noise at both levels. See Figure \ref{fig:Zoccolotti_letter_word} and the main text for details.}
\scalebox{0.75}{
\hspace{-2em}
\begin{tabular}{c||ll||ll||}
\cline{2-5}& \multicolumn{2}{c||}{\textsc{4 - letter word}} & \multicolumn{2}{c||}{\textsc{8 - letter word}}\\ 
\cline{2-5} & \multicolumn{1}{c|}{\textsc{Accuracy}} & \textsc{Probability} & \multicolumn{1}{|c|}{\textsc{Accuracy}} & \textsc{Probability} \\ \hline \hline
\multicolumn{1}{||c||}{\textsc{Control Model}} & \multicolumn{1}{c|}{$100$} & $1.00 \pm 1.03\cdot 10^{-14}$ & \multicolumn{1}{c|}{$100$} & $1.00 \pm7.85\cdot 10^{-15}$ \\ \hline
\multicolumn{1}{||c||}{\textsc{DM -Noise on Level 1}} & \multicolumn{1}{c|}{$100$} & $1.00 \pm 3.32\cdot 10^{-04}$ & \multicolumn{1}{c|}{$100$} & $0.99 \pm 3.82\cdot 10^{-04}$ \\ \hline
\multicolumn{1}{||c||}{\textsc{DM -Noise on Level 2}} & \multicolumn{1}{c|}{$99$} & $0.99 \pm 3.73\cdot 10^{-15}$ & \multicolumn{1}{c|}{97} & $0.97 \pm 1.38\cdot 10^{-02}$ \\ \hline
\multicolumn{1}{||c||}{\textsc{Dyslexic Model}} & \multicolumn{1}{c|}{$99$} & $0.99 \pm 3.32 \cdot 10^{-04}$ & \multicolumn{1}{c|}{$97$} & $0.97 \pm 1.15\cdot 10^{-02}$             \\ \hline
\end{tabular}
}
\label{tab:Letter_AccProb}
\end{table}

\begin{table}[ht!]
\caption{Simulation 1. T-tests for simulation of 4-letter and 8-letter word reading.
The table displays the results of T-tests conducted on the total number of saccades for different models with respect to the \emph{Control model} (CM): the \emph{Dyslexic model} (DM) variant with noise at Level 1 (syllable) only, the \emph{Dyslexic model} (DM) variant with noise at Level 2 (word) only, and the \emph{Dyslexic model} (DM) with noise at both levels. See Figure \ref{fig:Lett} and the main text for details.}
\resizebox{1 \textwidth}{!}{ 
\hspace{-2em}
\begin{tabular}{c||ll||}
\cline{2-3} 
& \multicolumn{2}{c||}{\textsc{Total number of saccades}} 
\\\cline{2-3}
& \multicolumn{1}{c|}{\textsc{4 - letter word}} & \multicolumn{1}{c||}{\textsc{8 - letter word}} \\ \hline \hline
\multicolumn{1}{||c||}{\textsc{DM - Noise Level 1}} & \multicolumn{1}{l|}{$F[1,198]=17.6,\mathbf{\vb*{p}\approx 4.1\cdot 10^{-05}}$} & $F[1,198]=12.7,\mathbf{\vb*{p}\approx 4.5\cdot 10^{-04}}$ \\ \hline
\multicolumn{1}{||c||}{\textsc{DM - Noise Level 2}} & \multicolumn{1}{l|}{$F[1,196]=4.4,\mathbf{\vb*{p}\approx 3.6\cdot 10^{-02}}$}   & $F[1,190]=5.2,\mathbf{\vb*{p}\approx 2.4\cdot 10^{-02}}$\\ \hline
\multicolumn{1}{||c||}{\textsc{Dyslexic Model}} & \multicolumn{1}{l|}{$F[1,196]=16.7,\mathbf{\vb*{p}\approx 6.4\cdot 10^{-05}}$}  & $F[1,192]=10.7,\mathbf{\vb*{p}\approx 1.3\cdot 10^{-03}}$
\\ \hline
\end{tabular}
}
\label{tab:ANOVA_Letter_sacc}
\end{table}
\begin{table}[ht!]
\caption{Simulation 1. T-tests for simulation of 4-letter and 8-letter word reading.
The table displays the results of T-tests conducted on the number of backward saccades for different models with respect to the \emph{Control model} (CM): the \emph{Dyslexic model} (DM) variant with noise at Level 1 (syllable) only, the \emph{Dyslexic model} (DM) variant with noise at Level 2 (word) only, and the \emph{Dyslexic model} (DM) with noise at both levels. See Figure \ref{fig:Lett} and the main text for details.}
\resizebox{1 \textwidth}{!}{ 
\hspace{-2em}
\begin{tabular}{c||ll||}
\cline{2-3} 
& \multicolumn{2}{c||}{\textsc{Number of backward saccades}}
\\\cline{2-3}
& \multicolumn{1}{c|}{\textsc{4 - letter word}} & \multicolumn{1}{c||}{\textsc{8 - letter word}} \\ \hline \hline
\multicolumn{1}{||c||}{\textsc{DM - Noise Level 1}} & \multicolumn{1}{l|}{$F[1,198]=21.5,\mathbf{\vb*{p}\approx 6.5\cdot 10^{-06}}$}   & $F[1,198]=17.5,\mathbf{\vb*{p}\approx 4.4\cdot 10^{-05}}$ \\ \hline
\multicolumn{1}{||c||}{\textsc{DM - Noise Level 2}} & \multicolumn{1}{l|}{$F[1,198]=4.8,\mathbf{\vb*{p}\approx 3.0\cdot 10^{-02}}$}   &  $F[1,190]=5.7,\mathbf{\vb*{p}\approx 1.8\cdot 10^{-02}}$ \\ \hline
\multicolumn{1}{||c||}{\textsc{Dyslexic Model}} & \multicolumn{1}{l|}{$F[1,196]=21.7,\mathbf{\vb*{p}\approx 5.8\cdot 10^{-06}}$} & $F[1,192]=13.3,\mathbf{\vb*{p}\approx 3.4\cdot 10^{-04}}$
\\ \hline
\end{tabular}
}
\label{tab:ANOVA_Letter_backsacc}
\end{table}

\begin{table}[ht!]
\caption{Simulation 1. T-tests for simulation of 4-letter and 8-letter word reading.
The table displays the results of T-tests conducted on the amplitude of forward saccades for different models with respect to the \emph{Control model} (CM): the \emph{Dyslexic model} (DM) variant with noise at Level 1 (syllable) only, the \emph{Dyslexic model} (DM) variant with noise at Level 2 (word) only, and the \emph{Dyslexic model} (DM) with noise at both levels. See Figure \ref{fig:Zoccolotti_letter_amplitude} and the main text for details.}
\resizebox{1 \textwidth}{!}{ 
\hspace{-2em}
\begin{tabular}{l||cc||}
\cline{2-3}
 &
  \multicolumn{2}{c||}{\textsc{Amplitude of forward   saccades}} \\ \cline{2-3} 
 &
  \multicolumn{1}{c|}{\textsc{4  - letter word}} &
  \textsc{8 - letter word} \\ \hline \hline
\multicolumn{1}{||c||}{\textsc{Dyslexic Model}} &
  \multicolumn{1}{c|}{$F[1,844]=9.4,\mathbf{\vb*{p}\approx 2.2\cdot 10^{-03}}$} &
  $F[1,678]=46.5,\mathbf{\vb*{p}\approx 2.0\cdot 10^{-11}}$ \\ \hline
\end{tabular}}
  \label{tab:ANOVA_Letter_Forw_AMPLITUDE}
\end{table}
\begin{table}[ht!]
\caption{Simulation 1. T-tests for simulation of 4-letter and 8-letter word reading.
The table displays the results of T-tests conducted on the amplitude of + backward saccades for different models with respect to the \emph{Control model} (CM): the \emph{Dyslexic model} (DM) variant with noise at Level 1 (syllable) only, the \emph{Dyslexic model} (DM) variant with noise at Level 2 (word) only, and the \emph{Dyslexic model} (DM) with noise at both levels. See Figure \ref{fig:Zoccolotti_letter_amplitude} and the main text for details.}
\resizebox{1 \textwidth}{!}{ 
\hspace{-2em}
\begin{tabular}{l||cc||}
\cline{2-3}
 &
  \multicolumn{2}{c||}{\textsc{Amplitude of backward   saccades}} \\ \cline{2-3} 
 &
  \multicolumn{1}{c|}{\textsc{4  - letter word}} &
  \textsc{8 - letter word} \\ \hline \hline
\multicolumn{1}{||c||}{\textsc{Dyslexic Model}} & \multicolumn{1}{c|}{$F[1,321]=133.5,\mathbf{\vb*{p}\approx 4.6\cdot 10^{-26}}$} &
  $F[1,312]=24.5,\mathbf{\vb*{p}\approx 1.2\cdot 10^{-06}}$ \\ \hline
\end{tabular}}
\label{tab:ANOVA_Letter_Backw_AMPLITUDE}
\end{table}

\subsubsection*{Supplementary tables for Simulation 2}

Table \ref{tab:Word_AccProb} display the accuracy and probability assigned to the correct words. Table \ref{tab:ANOVA_Word_Sacc} and Table \ref{tab:ANOVA_Word_backsacc} shows the T-tests on total saccades and back saccades. Table \ref{tab:ANOVA_Word_ForwAmpl} presents the T-tests for amplitude of forward and Table \ref{tab:ANOVA_Word_BackwAmpl} the T-test for amplitude of backward saccades respectively. The $p$-values lower to $0.05$ are indicated in bold.

\begin{table}[ht!]
\centering
\caption{Simulation 2: Reading sentences of 4 or 8 words. The table presents the accuracy and the probability assigned to the correct words by 4 different models: the \emph{Control model} (CM), the \emph{Dyslexic model} (DM) variant with noise at Level 1 only, the \emph{Dyslexic model} (DM) variant with noise at Level 2 only, the \emph{Dyslexic model} (DM) variant with noise at Level 3 only, and the \emph{Dyslexic model} (DM) with noise at all levels. See Figure \ref{fig:Zoccolotti_letter_word} and the main text for details.}
\resizebox{1 \textwidth}{!}{ 
\hspace{-2em}
\begin{tabular}{c||ll||ll||} \cline{2-5}& \multicolumn{2}  {c||}{\textsc{4 - word sentence}} & \multicolumn{2}{c||}{\textsc{8 - word sentence}}\\
\cline{2-5} & \multicolumn{1}{c|}{\textsc{Accuracy}} & \textsc{Probability} & \multicolumn{1}{c|}{\textsc{Accuracy}} & \textsc{Probability} \\ \hline \hline
\multicolumn{1}{||c||}{\textsc{Control Model}} & \multicolumn{1}{c|}{$100$} & $1.00 \pm 3.66 \cdot 10^{-16}$ & \multicolumn{1}{c|}{$100$} & $1.00 \pm 3.21 \cdot 10^{-12}$ \\ \hline
\multicolumn{1}{||c||}{\textsc{DM - Noise on Level 1}} & \multicolumn{1}{c|}{$100$} & $1.00 \pm 3.64\cdot 10^{-06}$ & \multicolumn{1}{c|}{$100$} & $1.00 \pm 5.69\cdot 10^{-08}$ \\ \hline
\multicolumn{1}{||c||}{\textsc{DM - Noise on Level 2}} & \multicolumn{1}{c|}{$100$} & $1.00 \pm 3.40 \cdot 10^{-07}$& \multicolumn{1}{c|}{$100$} & $1.00 \pm 3.21 \cdot 10^{-12}$ \\ \hline
\multicolumn{1}{||c||}{\textsc{DM - Noise on Level 3}} & \multicolumn{1}{c|}{$97$} & $0.97 \pm 4.25 \cdot 10^{-03}$ & \multicolumn{1}{c|}{$93$} & $0.93 \pm 1.04 \cdot 10^{-02}$ \\ \hline
\multicolumn{1}{||c||}{\textsc{Dyslexic Model}} & \multicolumn{1}{c|}{$96$} & $0.96 \pm 4.41 \cdot 10^{ -03}$ & \multicolumn{1}{c|}{$93$} & $0.93 \pm 1.04\cdot 10^{-02}$ \\ \hline
\end{tabular}}
\label{tab:Word_AccProb}
\end{table}

\begin{table}[ht!]
\caption{Simulation 2. T-tests for simulation of 4-word and 8-word sentence reading.
The table displays the results of T-tests conducted on the total number of saccades for different models with respect to the \emph{Control model} (CM): the \emph{Dyslexic model} (DM) variant with noise at Level 1 (syllable) only, the \emph{Dyslexic model} (DM) variant with noise at Level 2 (word) only,  the \emph{Dyslexic model} (DM) variant with noise at Level 3 (sentence) only, and the \emph{Dyslexic model} (DM) with noise at all levels. See Figure \ref{fig:Word} and the main text for details.}
\resizebox{1 \textwidth}{!}{\hspace{-2em}\begin{tabular}{c||ll||} 
\cline{2-3} 
& \multicolumn{2}{c||}{\textsc{Total number of saccades}}\\ 
\cline{2-3}
& \multicolumn{1}{c|}{\textsc{4 - word sentence}} & \multicolumn{1}{c||}{\textsc{8 - word sentence}} \\ \hline \hline
\multicolumn{1}{||c||}{\textsc{DM - Noise Level 1}} & \multicolumn{1}{l|}{$F[1,198]=0.02, ~ p\approx0.90$} & $F[1,198]=0.01, ~ p\approx0.92$ \\ \hline
\multicolumn{1}{||c||}{\textsc{DM - Noise Level 2}} & \multicolumn{1}{l|}{$F[1,198]=0.97, ~ p\approx0.45$} & $F[1,198]=0.45, ~~p\approx0.58$ \\ \hline
\multicolumn{1}{||c||}{\textsc{DM - Noise Level 3}} & \multicolumn{1}{l|}{$F[1,198]=66.8,\mathbf{\vb*{p}\approx 3.5\cdot 10^{-14}}$} & $F[1,196]=105.9, \mathbf{\vb*{p}\approx 3.9\cdot 10^{-20}}$ \\ \hline
\multicolumn{1}{||c||}{\textsc{Dyslexic Model}} & \multicolumn{1}{l|}{$F[1,198]=75.5, \mathbf{\vb*{p}\approx 1.4 \cdot 10^{-15}} $} &  $F[1,196]=113.6, \mathbf{\vb*{p}\approx 3.2\cdot 10^{-21}} $ \\ \hline
\end{tabular}}
\label{tab:ANOVA_Word_Sacc}
\end{table}

\begin{table}[ht!]
\caption{Simulation 2. T-tests for simulation of 4-word and 8-word sentence reading.
The table displays the results of T-tests conducted on the number of backward saccades for different models with respect to the \emph{Control model} (CM): the \emph{Dyslexic model} (DM) variant with noise at Level 1 (syllable) only, the \emph{Dyslexic model} (DM) variant with noise at Level 2 (word) only,  the \emph{Dyslexic model} (DM) variant with noise at Level 3 (sentence) only, and the \emph{Dyslexic model} (DM) with noise at all levels. See Figure \ref{fig:Word} and the main text for details.}
\resizebox{1 \textwidth}{!}{\hspace{-2em}\begin{tabular}{c||ll||} 
\cline{2-3} 
& \multicolumn{2}{c||}{\textsc{Number of backward saccades}} 
\\ 
\cline{2-3}
& \multicolumn{1}{c|}{\textsc{4 - word sentence}} & \multicolumn{1}{c||}{\textsc{8 - word sentence}} \\ \hline \hline
\multicolumn{1}{||c||}{\textsc{DM - Noise Level 1}} & \multicolumn{1}{l|}{$F[1,198]=0.45,~ p\approx0.52$} & $F[1,198]=0.82, ~ p\approx0.46$ \\ \hline
\multicolumn{1}{||c||}{\textsc{DM - Noise Level 2}} & \multicolumn{1}{l|}{$F[1,198]=0.01,~p\approx0.90$} & $F[1,198]=1.86, ~ p\approx0.17$ \\ \hline
\multicolumn{1}{||c||}{\textsc{DM - Noise Level 3}} & \multicolumn{1}{l|}{$F[1,198]=44.7, \mathbf{\vb*{p}\approx 2.3\cdot 10^{-10}}$}  & $F[1,196]=119.4, \mathbf{\vb*{p}\approx 5.1\cdot 10^{-22}}$ \\ \hline
\multicolumn{1}{||c||}{\textsc{Dyslexic Model}} & \multicolumn{1}{l|} {$F[1,198]=53.2,\mathbf{\vb*{p}\approx 7.1\cdot 10^{-12}}$} & $F[1,196]=117.7,\mathbf{\vb*{p}\approx 8.8\cdot 10^{-22}}$  \\ \hline
\end{tabular}}
\label{tab:ANOVA_Word_backsacc}
\end{table}


\begin{table}[ht!]
\caption{Simulation 2. T-tests for simulation of 4-word and 8-word sentence reading.
The table displays the results of T-tests conducted on the amplitude of forward saccades for different models with respect to the \emph{Control model} (CM): the \emph{Dyslexic model} (DM) variant with noise at Level 1 (syllable) only, the \emph{Dyslexic model} (DM) variant with noise at Level 2 (word) only, the \emph{Dyslexic model} (DM) variant with noise at Level 3 (sentence) only, and the \emph{Dyslexic model} (DM) with noise at all levels. See Figure \ref{fig:Zoccolotti_sentence_forward} and the main text for details.}
\resizebox{1 \textwidth}{!}{\hspace{-2em}\begin{tabular}{c||cc||}
\cline{2-3}
 &
  \multicolumn{2}{c||}{\textsc{Amplitude of forward saccades}} \\ \cline{2-3} 
 &
  \multicolumn{1}{c|}{\textsc{4 - word sentence}} &  \textsc{8 - word sentence} \\ \hline \hline
\multicolumn{1}{||c||}{\textsc{Dyslexic Model}} &
  \multicolumn{1}{c|}{$F[1,878]=18.0,\mathbf{\vb*{p}\approx 2.5\cdot 10^{-05}}$} &
  $F[1,1238]=10.9,\mathbf{\vb*{p}\approx 9.8\cdot 10^{-04}}$ \\ \hline
\end{tabular}}
\label{tab:ANOVA_Word_ForwAmpl}
\end{table}
\begin{table}[ht!]
\caption{Simulation 2. T-tests for simulation of 4-word and 8-word sentence reading.
The table displays the results of T-tests conducted on the amplitude of backward saccades for different models with respect to the \emph{Control model} (CM): the \emph{Dyslexic model} (DM) variant with noise at Level 1 (syllable) only, the \emph{Dyslexic model} (DM) variant with noise at Level 2 (word) only, the \emph{Dyslexic model} (DM) variant with noise at Level 3 (sentence) only, and the \emph{Dyslexic model} (DM) with noise at all levels. See Figure \ref{fig:Zoccolotti_sentence_forward} and the main text for details.}
\resizebox{1 \textwidth}{!}{\hspace{-2em}\begin{tabular}{c||cc||}
\cline{2-3}
 &
  \multicolumn{2}{c||}{\textsc{Amplitude of backward saccades}} \\ \cline{2-3} 
 &
  \multicolumn{1}{c|}{\textsc{4 - word sentence}} &
  \textsc{8 - word sentence} \\ \hline \hline
\multicolumn{1}{||c||}{\textsc{Dyslexic Model}} &
  \multicolumn{1}{c|}{$F[1,318]=14.7,\mathbf{\vb*{p}\approx 1.5\cdot 10^{-04}}$} &
  $F[1,350]=42.5,\mathbf{\vb*{p}\approx 2.5\cdot 10^{-10}}$  \\ \hline
\end{tabular}}
\label{tab:ANOVA_Word_BackwAmpl}
\end{table}

\subsubsection*{Simulation 4}
 Table \ref{tab:ANOVA_Letter_Forw_AMPLITUDE_SIM4} and Table \ref{tab:ANOVA_Letter_Backw_AMPLITUDE_SIM4} presents T-tests on total number of saccades and the number backward saccades when the model is augmented with the topic of the sentence at Level 3 of the hierarchy. The $p$-values lower to $0.05$ are indicated in bold.

\begin{table}[ht!]
\centering
\caption{Simulation 4. T-tests for simulation of a 9-word sentence reading, belonging to different topics.
The table displays the results of T-tests conducted on the total number of saccades for different models with respect to the \emph{Control model} (CM): the \emph{Dyslexic model} (DM) variant with noise at Level 1 (syllable) only, the \emph{Dyslexic model} (DM) variant with noise at Level 2 (word) only, the \emph{Dyslexic model} (DM) variant with noise at Level 3 (sentence) only, and the \emph{Dyslexic model} (DM) with noise at all levels. See Figure \ref{fig:Context} and the main text for details.}
\resizebox{0.9\textwidth}{!}{\hspace{0em}
\begin{tabular}{c||c||}
\cline{2-2}
\textbf{} & \textsc{Total number of saccades} \\ \hline \hline
\multicolumn{1}{||c||}{\textsc{Flat Priors VS Informative Priors}} & $F[1,198]=4.1,\mathbf{\vb*{p}\approx 4.1\cdot 10^{-03}}$ \\ \hline
\end{tabular}}
\label{tab:ANOVA_Letter_Forw_AMPLITUDE_SIM4}
\end{table}
\begin{table}[ht!]
\centering
\caption{Simulation 4. T-tests for simulation of a 9-word sentence reading, belonging to different topics.
The table displays the results of T-tests conducted on the number of backward saccades for different models with respect to the \emph{Control model} (CM): the \emph{Dyslexic model} (DM) variant with noise at Level 1 (syllable) only, the \emph{Dyslexic model} (DM) variant with noise at Level 2 (word) only, the \emph{Dyslexic model} (DM) variant with noise at Level 3 (sentence) only, and the \emph{Dyslexic model} (DM) with noise at all levels. See Figure \ref{fig:Context} and the main text for details.}
\resizebox{0.9\textwidth}{!}{\hspace{0em}
\begin{tabular}{c||c||}
\cline{2-2}
\textbf{} & \textsc{Number of backward saccades} \\ \hline \hline
\multicolumn{1}{||c||}{\textsc{Flat Priors VS Informative Priors}} & $F[1,198]=2.6,~~p \approx 0.11$ \\ \hline
\end{tabular}}
\label{tab:ANOVA_Letter_Backw_AMPLITUDE_SIM4}
\end{table}

\subsection*{Datasets}

In this section, we present the datasets utilized in our simulations. Information regarding content length in the datasets can be found in Table \ref{tab:Dictionary_characterLenght}. Furthermore, Table \ref{tab:Dictionary_simulation1} displays the dictionary of words used in Simulation 1, while Table \ref{tab:Dictionary_simulation2} provides insight into the sentences used in Simulation 2. Additionally, Simulation 4's sentences and corresponding labels are showcased in Table \ref{tab:Dictionary_simulation4}.

\begin{table}[ht!]
\centering
\caption{This table reports the mean number of characters, the standard deviation, minimum and maximum number of characters that compose the relative set of sentences used in the simulations.}
\scalebox{0.9}{
\begin{tabular}{||c|c|c|c|c||}
\cline{2-5}
\multicolumn{1}{l||}{} & \textsc{Mean} & \textsc{\begin{tabular}[c]{@{}c@{}}Standard\\ Deviation\end{tabular}} & \textsc{Min} & \textsc{Max} \\ \hline \hline
\multicolumn{1}{||c||}{\textsc{Simulation 2  (4 words)}}  & $25.96$ & $1.96$ & $21$ & $31$ \\ \hline \hline
\multicolumn{1}{||c||}{\textsc{Simulation 2   (8 words)}} & $50.24$ & $1.64$ & $47$ & $55$ \\ \hline \hline
\multicolumn{1}{||c||}{\textsc{Simulation 3  (4 words - 1 Unknown)}}  & $25.46$ & $1.92$ & $22$ & $32$ \\ \hline \hline
\multicolumn{1}{||c||}{\textsc{Simulation 4}}                      & $69.55$ & $5.02$ & $61$ & $82$ \\ \hline \hline
\end{tabular}}
\label{tab:Dictionary_characterLenght}
\end{table}

\newgeometry{top=1cm, bottom=1.5cm, left=0.5cm, right=0.5cm}
\footnotesize
\begin{longtable}{c||c||c||c||c||c||c||c||}
\caption{\footnotesize Dictionary used in Simulation 1  (words extracted from BERT Dictionary)} \\
\hline \hline
\multicolumn{1}{||c||}{\textsc{1 - letter}} &
  \textsc{2- letter} &
  \textsc{3- letter} &
  \textsc{4- letter} &
  \textsc{5- letter} &
  \textsc{6- letter} &
  \textsc{7- letter} &
  \textsc{8- letter} \\ \hline \hline
\multicolumn{1}{||c||}{A} & AC & AMC & ARES & AGREE & ACTING & AIRPORT & ACQUIRED \\ \hline
\multicolumn{1}{||c||}{B} & AL & ARM & ARTS & ANZAC & ADVENT & ALCOHOL & ACTUALLY \\ \hline
\multicolumn{1}{||c||}{C} & AM & ATA & ASKS & ARDEN & AIRMEN & ANGRILY & ADVANCED \\ \hline
\multicolumn{1}{||c||}{D} & AR & AUS & BEDS & ARIEL & ANGLIA & ARCADIA & ALPHABET \\ \hline
\multicolumn{1}{||c||}{E} & AW & BAO & BORE & AZURE & ARMAND & ASSURED & AMERICAN \\ \hline
\multicolumn{1}{||c||}{F} & AX & BEG & BRET & BARED & AUSTEN & AVENUES & APPLETON \\ \hline
\multicolumn{1}{||c||}{G} & BF & BEN & BUSY & BEGUN & BECKER & BALANCE & ARGUMENT \\ \hline
\multicolumn{1}{||c||}{H} & BI & BIT & BUTT & BIRDS & BEIRUT & BENGALS & ARRANGED \\ \hline
\multicolumn{1}{||c||}{I} & BK & BOP & CALE & BLAND & BERMAN & BERWICK & ATLANTIC \\ \hline
\multicolumn{1}{||c||}{J} & BO & BUY & CARL & BUICK & BIGGER & BOUNCED & BALANCED \\ \hline
\multicolumn{1}{||c||}{K} & BR & CAB & COAL & BULGE & BISHOP & BURUNDI & BASILICA \\ \hline
\multicolumn{1}{||c||}{L} & BT & CAD & COLA & BUTCH & BUMPER & CAREERS & BENEDICT \\ \hline
\multicolumn{1}{||c||}{M} & CA & CAI & COOL & CELLO & BURDEN & CEILING & BOTSWANA \\ \hline
\multicolumn{1}{||c||}{N} & CB & CEO & CUBS & CLEAR & CANTOR & CHANCEL & BOULDERS \\ \hline
\multicolumn{1}{||c||}{O} & CH & COE & DIRT & CLUBS & CAUSAL & CLICKED & BREACHED \\ \hline
\multicolumn{1}{||c||}{P} & CL & CPU & DOIN & CODES & CEREAL & CLIENTS & BREAKING \\ \hline
\multicolumn{1}{||c||}{Q} & CM & DEX & DOVE & COSTS & CESARE & COMMITS & CADILLAC \\ \hline
\multicolumn{1}{||c||}{R} & CN & DIG & DRIP & CRIED & CESSNA & CROOKED & CAMPBELL \\ \hline
\multicolumn{1}{||c||}{S} & CO & DIP & DRUM & CURLY & CHICKS & DERRICK & CARRYING \\ \hline
\multicolumn{1}{||c||}{T} & CP & EKI & ELLE & DANNY & CLOCKS & DESCENT & CATALINA \\ \hline
\multicolumn{1}{||c||}{U} & CU & EPA & EMIR & DATED & COINED & DESPAIR & COLISEUM \\ \hline
\multicolumn{1}{||c||}{V} & ED & ETA & EMMA & DEBUT & DECKER & DEVISED & COLUMBIA \\ \hline
\multicolumn{1}{||c||}{W} & EH & FIG & FANS & DISCS & DISMAY & EDUARDO & CONNECTS \\ \hline
\multicolumn{1}{||c||}{X} & EL & FLU & FEES & DOUBT & DRINKS & EMBASSY & CONVINCE \\ \hline
\multicolumn{1}{||c||}{Y} & EM & FUN & FIJI & DUCKS & EXEMPT & EMERSON & CUMMINGS \\ \hline
\multicolumn{1}{||c||}{Z} & EU & HAN & FINS & DUMMY & FACADE & EMPLOYS & CURRENCY \\ \hline
                        & FI & HBO & FLOP & ECOLE & FINLEY & ENEMIES & CYCLONES \\ \cline{2-8} 
                        & FL & HEM & FLUX & ELIZA & GENTRY & ESCAPES & CYLINDER \\ \cline{2-8} 
                        & FT & HIT & FRAN & ESSEN & HAILEY & EXHAUST & DARKENED \\ \cline{2-8} 
                        & FU & HOP & GIFT & FEMME & HANNAH & FAINTLY & DECEASED \\ \cline{2-8} 
                        & FX & HUE & GIGS & FISTS & HASSAN & FLEDGED & DECIDING \\ \cline{2-8} 
                        & GE & IBN & GLOW & FLOOR & HATRED & FORSTER & DISABLED \\ \cline{2-8} 
                        & GI & ICE & GRAM & FLOYD & HELMUT & FREEWAY & DISASTER \\ \cline{2-8} 
                        & GO & ICH & GREY & GOUGH & HOOVER & FRONTED & DONNELLY \\ \cline{2-8} 
                        & GS & ICT & HAAS & GROWS & HUTTON & GALILEO & EMERITUS \\ \cline{2-8} 
                        & GT & ILE & HANK & GRUNT & HYBRID & GEOLOGY & EMPLOYER \\ \cline{2-8} 
                        & GU & ION & HILL & GUARD & INCOME & GLIDING & EPILOGUE \\ \cline{2-8} 
                        & HA & IRA & HISS & HELLO & INPUTS & HAMBURG & EREBIDAE \\ \cline{2-8} 
                        & HC & KIM & HUEY & HIRES & INTERN & HANDLED & EVACUATE \\ \cline{2-8} 
                        & HM & LAL & INDY & HORDE & JOSEPH & HIGHEST & EVOLVING \\ \cline{2-8} 
                        & HO & LAY & JACE & JOINT & JUAREZ & HONNEUR & EXAMINER \\ \cline{2-8} 
                        & HU & LEG & JAIN & JOYCE & KHYBER & HORNETS & FABULOUS \\ \cline{2-8} 
                        & IK & LES & JEAN & LATER & LANDON & HOWEVER & FIGHTERS \\ \cline{2-8} 
                        & IN & LOU & JING & LEANS & LENGTH & HOWLING & FRANKLIN \\ \cline{2-8} 
                        & IO & LTD & KRIS & LEAVE & LENNON & HUNCHED & FRICTION \\ \cline{2-8} 
                        & IR & MED & LAMA & LOOSE & LIKELY & INHALED & GESTURES \\ \cline{2-8} 
                        & IX & MPH & LANG & LOVER & LILITH & IRANIAN & GROUPING \\ \cline{2-8} 
                        & JA & MPS & LETO & LUCHA & LITTER & JEALOUS & HOMICIDE \\ \cline{2-8} 
                        & JD & MSC & LIAR & LYDIA & MADRAS & JUSTINE & IMPERIAL \\ \cline{2-8} 
                        & KE & MUD & MACK & MAHAL & MANUEL & KNIGHTS & INCURRED \\ \cline{2-8} 
                        & KG & NBA & MARS & MERGE & MARGIN & LABELED & INITIATE \\ \cline{2-8} 
                        & KN & NHL & MASK & MINSK & MARKED & LASTING & INJURIES \\ \cline{2-8} 
                        & KO & NHS & MEAT & MISSY & MCLEAN & LEBANON & INSANITY \\ \cline{2-8} 
                        & KS & NIK & MIKA & MIXED & MOLINA & MANNING & JUDICIAL \\ \cline{2-8} 
                        & LC & NOS & MOJO & MOMMA & MUSEUM & MCBRIDE & LABRADOR \\ \cline{2-8}  
                        & LI & NOV & MOOD & NAILS & MYSELF & MORALLY & MAGAZINE \\ \cline{2-8} 
                        & LP & NPR & MUCH & NORMA & NEWELL & MUSCLED & MARATHON \\ \cline{2-8} 
                        & MG & NYC & NEAT & OLDER & ONIONS & NATALIE & MEANINGS \\ \cline{2-8} 
                        & MI & OFF & NINO & OUTTA & OPENLY & NEURONS & MOHAMMAD \\ \cline{2-8} 
                        & ML & OUT & NOSE & PAPUA & OPPOSE & NEVILLE & MONSIEUR \\ \cline{2-8} 
                        & MR & PAT & NRHP & PATIO & PANZER & NIKOLAI & MOROCCAN \\ \cline{2-8} 
                        & NA & PAU & OBOE & PIANO & PAVING & NILSSON & MOUNTING \\ \cline{2-8} 
                        & NH & PBA & OWLS & PILES & PAYTON & NOMINAL & MUSHROOM \\ \cline{2-8} 
                        & NI & PBS & PASS & PLUTO & PETALS & OFFENSE & OLYMPIAD \\ \cline{2-8} 
                        & NO & PCS & PERU & PROSE & PILOTS & ONSTAGE & PACKAGES \\ \cline{2-8} 
                        & NS & PEI & PORN & QAEDA & PLANES & PATCHES & PATERSON \\ \cline{2-8} 
                        & NT & PEW & RAAF & RALLY & POLITE & PEPTIDE & PHILLIPS \\ \cline{2-8} 
                        & NZ & PHI & RAMP & REEFS & POTION & PERFECT & POPULACE \\ \cline{2-8} 
                        & OG & PHP & RAMS & REVUE & PRAGUE & PEUGEOT & POUNDING \\ \cline{2-8} 
                        & OH & POD & RUDD & ROACH & PRICED & PIONEER & PREACHER \\ \cline{2-8} 
                        & OL & PUN & SANK & ROUEN & PULSES & PLANNED & PRESTIGE \\ \cline{2-8} 
                        & OP & ROE & SAXE & RUSSO & PUPILS & PLAYFUL & QUARTERS \\ \cline{2-8} 
                        & OS & ROM & SITU & SCOTT & RASHID & PORTICO & RAILROAD \\ \cline{2-8} 
                        & OZ & RUM & SIZE & SHORE & REPORT & PROTEST & RECEIVES \\ \cline{2-8} 
                        & PH & RYE & SLID & SINGH & RIPLEY & RAPIDLY & REJOINED \\ \cline{2-8} 
                        & PO & SAO & SODA & SITED & ROBERT & READERS & SCENARIO \\ \cline{2-8} 
                        & PR & SEA & SOON & SLAMS & RUNNER & RECITAL & SEMINOLE \\ \cline{2-8} 
                        & RC & SHI & STAY & SPRAY & SAILOR & REDWOOD & SHOOTOUT \\ \cline{2-8} 
                        & RR & SHU & STYX & SQUAT & SAXONY & REISSUE & SHOWERED \\ \cline{2-8} 
                        & RY & SPP & SUFI & STADE & SCARES & ROOSTER & SINGULAR \\ \cline{2-8} 
                        & SE & SSR & SWAM & STAFF & SCREAM & RUINING & SOMERSET \\ \cline{2-8} 
                        & SF & SUE & TEAM & STALL & SEIZED & SECURED & SOUTHEND \\ \cline{2-8} 
                        & SM & TAI & THRU & STATE & SESAME & SILENCE & SPECIALS \\ \cline{2-8} 
                        & SQ & TEA & TODD & STERN & SEVERN & SMASHED & SPEEDING \\ \cline{2-8} 
                        & SR & TEX & TOLD & STORM & SIGNAL & SPELLED & STRANDED \\ \cline{2-8} 
                        & TC & TIS & TOPS & STUMP & SPEAKS & STATION & STRIPPED \\ \cline{2-8} 
                        & TU & TNA & TRIP & TOUGH & SPINES & STEWART & SYMBOLIC \\ \cline{2-8} 
                        & TV & UFO & TUNA & TRANS & STALLS & STRANGE & TARGETED \\ \cline{2-8} 
                        & TX & VAR & UGLY & TREAT & STARTS & STRIKES & TAXATION \\ \cline{2-8} 
                        & UP & VII & USER & TREES & SURFER & SWELLED & THROTTLE \\ \cline{2-8} 
                        & UR & WAT & VEGA & VAULT & TORINO & TIGHTLY & TIMELINE \\ \cline{2-8} 
                        & VA & WAY & VIDA & VIJAY & TREVOR & TOPICAL & TOURISTS \\ \cline{2-8} 
                        & VE & WOW & VISA & VISAS & TURTLE & TURTLES & UNCOMMON \\ \cline{2-8} 
                        & VP & XII & WEEP & VOMIT & VENDOR & ULYSSES & VALENTIN \\ \cline{2-8} 
                        & VU & XVI & WHOA & WEARS & VERBAL & UNITING & VIGOROUS \\ \cline{2-8} 
                        & WC & YES & WIFE & WHORE & WANTED & UNNAMED & VILLAINS \\ \cline{2-8} 
                        & XI & YET & WINS & WRAPS & WASHED & VACANCY & WARRIORS \\ \cline{2-8} 
                        & YA & YOO & WITS & WRIST & WIRING & VISIBLE & WEREWOLF \\ \cline{2-8} 
                        & YE & ZEE & YUKI & YEMEN & WORTHY & WALKING & WORRYING \\ \cline{2-8} 
                        & YO & ZEV & ZOOM & YOUTH & XAVIER & WARTIME & ZIMBABWE \\ \cline{2-8}
\end{longtable}
\label{tab:Dictionary_simulation1}
\restoregeometry

\newgeometry{top=1cm, bottom=1.5cm, left=0.5cm, right=0.5cm}
\begin{small}
\footnotesize
\begin{longtable}{||p{5cm}||p{12cm}||}
\caption{\footnotesize Dictionary used in Simulation 2}  \\
\hline
\hline
FAITH CONQUERS ALL OBSTACLES      & FAITH CONQUERS ALL OBSTACLES AND TRASCENDS ALL ADVERSITY       \\ \hline
FAITH CONQUERS EACH HURDLE        & FAITH CONQUERS ALL OBSTACLES AND TRASCENDS ALL MISFORTUNE      \\ \hline
FAITH CONQUERS EACH SETBACK       & FAITH CONQUERS EACH HURDLE BUT TRASCENDS ALL MISFORTUNE        \\ \hline
FAITH CONQUERS EVERY CHALLENGES   & FAITH CONQUERS EACH SETBACK BUT SURMOUNTS ALL OBSTACLES        \\ \hline
FAITH CONQUERS EVERY OBSTACLES    & FAITH CONQUERS EVERY OBSTACLES AND PREVAILS OVER ADVERSITY     \\ \hline
FAITH DEFEATS ALL OBSTACLES       & FAITH CONQUERS EVERY OBSTACLES AND SURMOUNTS ALL OBSTACLES     \\ \hline
FAITH OVERCOMES ALL HINDRANCES    & FAITH DEFEATS ALL OBSTACLES AND TRASCENDS ALL ADVERSITY        \\ \hline
FAITH OVERCOMES ALL OBSTACLES     & FAITH DEFEATS ALL OBSTACLES AND TRASCENDS ALL MISFORTUNE       \\ \hline
FAITH OVERPOWERS ALL LIMITATIONS  & FAITH OVERCOMES ALL HINDRANCES AND TRASCENDS OVER MISFORTUNE   \\ \hline
FAITH PREVAILS AGAINST BARRIERS   & FAITH OVERCOMES ALL OBSTACLES AND TRASCENDS ALL ADVERSITY      \\ \hline
FAITH PREVAILS AGAINST OBSTACLES  & FAITH OVERCOMES ALL OBSTACLES AND TRASCENDS ALL MISFORTUNE     \\ \hline
FAITH PREVAILS OVER EVERY ADVERSITY & FAITH OVERPOWERS ALL LIMITATIONS BUT TRASCENDS ALL MISFORTUNE \\ \hline
FAITH PREVAILS OVER OBSTACLES     & FAITH PREVAILS AGAINST BARRIERS AND TRASCENDS OVER MISFORTUNE  \\ \hline
FAITH SURMOUNTS ALL OBSTACLES     & FAITH PREVAILS AGAINST OBSTACLES AND TRASCENDS ALL MISFORTUNE  \\ \hline
FAITH SURPASSES ALL OBSTACLES     & FAITH PREVAILS EVERY ADVERSITY BUT SURMOUNTS ALL OBSTACLES     \\ \hline
FAITH SURPASSES ALL OPPOSITION    & FAITH PREVAILS OVER ADVERSITY BUT SURMOUNTS ALL OBSTACLES      \\ \hline
FAITH TRIUMPHS AGAINST OBSTACLES  & FAITH PREVAILS OVER OBSTACLES AND TRASCENDS ALL MISFORTUNE     \\ \hline
FAITH TRIUMPHS OVER CHALLENGES    & FAITH SURMOUNTS ALL OBSTACLES AND SURMOUNTS ALL MISFORTUNE     \\ \hline
FAITH TRIUMPHS OVER OBSTACLES     & FAITH SURPASSES ALL OBSTACLES AND TRASCENDS ALL ADVERSITY      \\ \hline
FAITH VANQUISHES ALL OBSTACLES    & FAITH SURPASSES ALL OBSTACLES AND TRASCENDS ALL MISFORTUNE     \\ \hline
FAITH VANQUISHES EVERY DIFFICULTY & FAITH SURPASSES ALL OBSTACLES BUT SURMOUNTS ALL ADVERSITY      \\ \hline
HOPE CONQUERS ALL LIMITATIONS     & FAITH SURPASSES ALL OPPOSITION BUT SURMOUNTS ALL MISFORTUNE    \\ \hline
HOPE CONQUERS ALL OBSTACLES       & FAITH TRIUMPHS AGAINST OBSTACLES AND TRASCENDS ALL MISFORTUNE  \\ \hline
HOPE CONQUERS EACH CHALLENGES     & FAITH TRIUMPHS OVER CHALLENGES AND TRASCENDS OVER MISFORTUNE   \\ \hline
HOPE CONQUERS EACH HINDRANCES     & FAITH TRIUMPHS OVER OBSTACLES AND TRASCENDS ALL ADVERSITY      \\ \hline
HOPE CONQUERS EACH HURDLE         & FAITH TRIUMPHS OVER OBSTACLES AND TRASCENDS ALL MISFORTUNE     \\ \hline
HOPE CONQUERS EACH INHIBITION     & FAITH VANQUISHES ALL OBSTACLES AND TRASCENDS ALL MISFORTUNE    \\ \hline
HOPE CONQUERS EACH LIMITATIONS    & FAITH VANQUISHES EVERY DIFFICULTY BUT TRASCENDS ALL MISFORTUNE \\ \hline
HOPE CONQUERS EACH SETBACK        & HOPE CONQUERS ALL OBSTACLES AND TRASCENDS ALL ADVERSITY        \\ \hline
HOPE CONQUERS EVERY ADVERSITY     & HOPE CONQUERS ALL OBSTACLES AND TRASCENDS ALL MISFORTUNE       \\ \hline
HOPE CONQUERS EVERY BARRIERS      & HOPE CONQUERS EACH SETBACK BUT SURMOUNTS ALL OBSTACLES         \\ \hline
HOPE CONQUERS EVERY CHALLENGES    & HOPE DEFEATS ALL OBSTACLES AND TRASCENDS ALL ADVERSITY         \\ \hline
HOPE CONQUERS EVERY DIFFICULTY    & HOPE DEFEATS ALL OBSTACLES AND TRASCENDS ALL MISFORTUNE        \\ \hline
HOPE CONQUERS EVERY HURDLE        & HOPE OVERCOMES ALL OBSTACLES AND TRASCENDS ALL ADVERSITY       \\ \hline
HOPE CONQUERS EVERY OBSTACLES     & HOPE OVERCOMES ALL OBSTACLES AND TRASCENDS ALL MISFORTUNE      \\ \hline
HOPE CONQUERS EVERY RESTRAINT     & HOPE PREVAILS OVER OBSTACLES AND TRASCENDS ALL ADVERSITY       \\ \hline
HOPE DEFEATS ALL OPPOSITION       & HOPE PREVAILS OVER OBSTACLES AND TRASCENDS ALL MISFORTUNE      \\ \hline
HOPE OVERCOMES ALL CHALLENGES     & HOPE SURPASSES ALL OBSTACLES AND TRASCENDS ALL ADVERSITY       \\ \hline
HOPE OVERCOMES EACH BOUNDARY      & HOPE SURPASSES ALL OBSTACLES AND TRASCENDS ALL MISFORTUNE      \\ \hline
HOPE OVERCOMES EACH HINDRANCES    & HOPE TRIUMPHS OVER OBSTACLES AND TRASCENDS ALL ADVERSITY       \\ \hline
HOPE PREVAILS AGAINST ADVERSITY   & HOPE TRIUMPHS OVER OBSTACLES AND TRASCENDS ALL MISFORTUNE      \\ \hline
HOPE PREVAILS AGAINST BARRIERS    & LOVE CONQUERS ALL ADVERSITY AND TRASCENDS ALL ADVERSITY        \\ \hline
HOPE PREVAILS AGAINST IMPEDIMENT  & LOVE CONQUERS ALL ADVERSITY AND TRASCENDS ALL MISFORTUNE       \\ \hline
HOPE PREVAILS OVER OBSTACLES      & LOVE CONQUERS ALL ADVERSITY BUT TRASCENDS ALL MISFORTUNE       \\ \hline
HOPE SURMOUNTS ALL DIFFICULTIES   & LOVE CONQUERS ALL ADVERSITY BUT TRASCENDS EVERY MISFORTUNE     \\ \hline
HOPE SURPASSES ALL OBSTACLES      & LOVE CONQUERS ALL BOUNDARIES AND TRASCENDS ALL ADVERSITY       \\ \hline
HOPE SURPASSES EVERY CONFINEMENT  & LOVE CONQUERS ALL BOUNDARIES AND TRASCENDS ALL MISFORTUNE      \\ \hline
HOPE SURPASSES EVERY OBSTACLES    & LOVE CONQUERS ALL BOUNDARIES BUT TRASCENDS ALL MISFORTUNE      \\ \hline
HOPE TRANSCENDS EACH RESTRICTION  & LOVE CONQUERS ALL BOUNDARIES BUT TRASCENDS EVERY MISFORTUNE    \\ \hline
HOPE TRIUMPHS AGAINST OBSTACLES   & LOVE CONQUERS ALL DIFFERENCES AND TRASCENDS ALL ADVERSITY      \\ \hline
HOPE TRIUMPHS OVER CHALLENGES     & LOVE CONQUERS ALL DIFFERENCES AND TRASCENDS ALL MISFORTUNE     \\ \hline
HOPE VANQUISHES ALL HINDRANCES    & LOVE CONQUERS ALL DIFFERENCES BUT TRASCENDS ALL MISFORTUNE     \\ \hline
LOVE OVERCOMES EACH HINDRANCES    & LOVE CONQUERS ALL DIFFERENCES BUT TRASCENDS EVERY MISFORTUNE   \\ \hline
LOVE CONQUERS ALL ADVERSITY       & LOVE CONQUERS ALL OBSTACLES AND TRASCENDS ALL ADVERSITY        \\ \hline
LOVE CONQUERS ALL BOUNDARIES      & LOVE CONQUERS ALL OBSTACLES AND TRASCENDS ALL MISFORTUNE       \\ \hline
LOVE CONQUERS ALL DIFFERENCES     & LOVE CONQUERS ALL OBSTACLES BUT TRASCENDS ALL ADVERSITY        \\ \hline
LOVE CONQUERS ALL OBSTACLES       & LOVE CONQUERS ALL OBSTACLES BUT TRASCENDS EVERY ADVERSITY      \\ \hline
LOVE CONQUERS ANY OBSTACLES       & LOVE CONQUERS EVERY BARRIERS AND PREVAILS OVER ADVERSITY       \\ \hline
LOVE CONQUERS EACH OBSTACLES      & LOVE CONQUERS EVERY OBSTACLES BUT TRASCENDS ALL MISFORTUNE     \\ \hline
LOVE CONQUERS EVERY BARRIERS      & LOVE DEFEATS ALL OBSTACLES AND PREVAILS OVER ADVERSITY         \\ \hline
LOVE CONQUERS EVERY CHALLENGES    & LOVE OVERCOMES ALL OBSTACLES AND CONQUERS ALL ADVERSITY        \\ \hline
LOVE CONQUERS EVERY CONFINEMENT   & LOVE OVERCOMES ALL OBSTACLES AND CONQUERS ALL MISFORTUNE       \\ \hline
LOVE CONQUERS EVERY DIFFICULTY    & LOVE OVERCOMES ALL OBSTACLES AND CONQUERS EVERY OBSTACLES      \\ \hline
LOVE CONQUERS EVERY HINDRANCES    & LOVE OVERCOMES ALL OBSTACLES AND TRASCENDS ALL ADVERSITY       \\ \hline
LOVE CONQUERS EVERY HURDLE        & LOVE OVERCOMES ALL OBSTACLES AND TRASCENDS ALL MISFORTUNE      \\ \hline
LOVE CONQUERS EVERY LIMITATIONS   & LOVE OVERCOMES ALL OBSTACLES BUT CONQUERS ALL ADVERSITY        \\ \hline
LOVE CONQUERS EVERY OBSTACLES     & LOVE OVERCOMES ALL OBSTACLES BUT CONQUERS EVERY ADVERSITY      \\ \hline
LOVE CONQUERS EVERY SETBACK       & LOVE OVERCOMES ALL OBSTACLES BUT TRASCENDS ALL ADVERSITY       \\ \hline
LOVE DEFEATS ALL OBSTACLES        & LOVE OVERCOMES ALL OBSTACLES BUT TRASCENDS EACH ADVERSITY      \\ \hline
LOVE DEFEATS ALL OPPOSITION       & LOVE OVERCOMES ALL OBSTACLES BUT TRASCENDS EVERY ADVERSITY     \\ \hline
LOVE OVERCOMES ALL OBSTACLES      & LOVE PREVAILS AGAINST CHALLENGES AND PREVAILS OVER ADVERSITY   \\ \hline
LOVE PREVAILS AGAINST CHALLENGES    & LOVE PREVAILS AGAINST OBSTACLES AND OVERCOMES OVER MISFORTUNE \\ \hline
LOVE PREVAILS AGAINST OBSTACLES   & LOVE PREVAILS AGAINST OBSTACLES AND PREVAILS OVER ADVERSITY    \\ \hline
LOVE PREVAILS OVER OBSTACLES      & LOVE PREVAILS OVER OBSTACLES AND TRASCENDS ALL MISFORTUNE      \\ \hline
LOVE SURMOUNTS ALL OBSTACLES      & LOVE SURMOUNTS ALL OBSTACLES BUT TRASCENDS ALL ADVERSITY       \\ \hline
LOVE SURPASSES ALL OBSTACLES      & LOVE SURMOUNTS ALL OBSTACLES BUT TRASCENDS ALL MISFORTUNE      \\ \hline
LOVE TRIUMPHS AGAINST OBSTACLES   & LOVE SURPASSES ALL OBSTACLES AND PREVAILS OVER ADVERSITY       \\ \hline
LOVE TRIUMPHS OVER OBSTACLES      & LOVE TRIUMPHS AGAINST OBSTACLES AND DEFEATS ALL ADVERSITY      \\ \hline
LOVE VANQUISHES ALL OBSTACLES     & LOVE TRIUMPHS AGAINST OBSTACLES AND DEFEATS ALL OBSTACLES      \\ \hline
UNITY BREAKS THROUGH ALL BARRIERS & LOVE TRIUMPHS AGAINST OBSTACLES AND DEFEATS EACH ADVERSITY     \\ \hline
UNITY CONQUERS AGAINST ODDS       & LOVE TRIUMPHS AGAINST OBSTACLES AND DEFEATS OVER ADVERSITY     \\ \hline
UNITY CONQUERS ALL OBSTACLES      & LOVE TRIUMPHS AGAINST OBSTACLES AND DEFEATS OVER OBSTACLES     \\ \hline
UNITY CONQUERS DESPITE ODDS       & LOVE TRIUMPHS AGAINST OBSTACLES BUT DEFEATS ALL ADVERSITY      \\ \hline
UNITY DEFEATS AGAINST ODDS        & LOVE TRIUMPHS AGAINST OBSTACLES BUT DEFEATS ALL OBSTACLES      \\ \hline
UNITY DEFEATS ALL OPPOSITION      & LOVE TRIUMPHS AGAINST OBSTACLES BUT DEFEATS EACH ADVERSITY     \\ \hline
UNITY OVERCOMES AGAINST ODDS      & LOVE TRIUMPHS AGAINST OBSTACLES BUT DEFEATS OVER ADVERSITY     \\ \hline
UNITY OVERCOMES ALL CHALLENGES    & LOVE TRIUMPHS AGAINST OBSTACLES BUT DEFEATS OVER OBSTACLES     \\ \hline
UNITY OVERPOWERS ALL RESISTANCE   & LOVE TRIUMPHS OVER OBSTACLES AND PREVAILS OVER ADVERSITY       \\ \hline
UNITY PREVAILS AGAINST ODDS       & UNITY CONQUERS ALL ADVERSITY AND TRASCENDS ALL MISFORTUNE      \\ \hline
UNITY PREVAILS AMIDST ODDS        & UNITY CONQUERS ALL ADVERSITY AND TRASCENDS EACH MISFORTUNE     \\ \hline
UNITY PREVAILS DESPITE ODDS       & UNITY CONQUERS ALL BOUNDARIES AND TRASCENDS ALL ADVERSITY      \\ \hline
UNITY PREVAILS OVER ODDS          & UNITY CONQUERS ALL BOUNDARIES AND TRASCENDS ALL MISFORTUNE     \\ \hline
UNITY RISES ABOVE ALL OBSTACLES   & UNITY CONQUERS ALL BOUNDARIES AND TRASCENDS EACH ADVERSITY     \\ \hline
UNITY SUBDUES ALL DIFFICULTIES    & UNITY CONQUERS ALL BOUNDARIES AND TRASCENDS EACH MISFORTUNE    \\ \hline
UNITY SUCCEEDS AGAINST ODDS       & UNITY CONQUERS ALL DIFFERENCES AND TRASCENDS ALL MISFORTUNE    \\ \hline
UNITY SURMOUNTS ALL OBSTACLES     & UNITY CONQUERS ALL OBSTACLES AND TRASCENDS ALL MISFORTUNE      \\ \hline
UNITY SURPASSES AGAINST ODDS      & UNITY OVERCOMES ALL OBSTACLES AND CONQUERS ALL ADVERSITY       \\ \hline
UNITY TRIUMPHS AGAINST ODDS       & UNITY OVERCOMES ALL OBSTACLES AND CONQUERS ALL MISFORTUNE      \\ \hline
UNITY TRIUMPHS OVER BARRIERS      & UNITY OVERCOMES ALL OBSTACLES AND TRASCENDS ALL ADVERSITY      \\ \hline
UNITY VANQUISHES ALL HINDRANCES   & UNITY OVERCOMES ALL OBSTACLES AND TRASCENDS ALL MISFORTUNE     \\ \hline \hline
\end{longtable}
\label{tab:Dictionary_simulation2}
\end{small}

\textcolor{rev}{
\centering
\newgeometry{top=1cm, bottom=1.5cm, left=0.5cm, right=0.5cm}
\begin{small}
\footnotesize
\begin{longtable}{||c||c||c||c||c||}
\caption{\footnotesize \textcolor{rev}{Dictionary used in Simulation 3 (new unknown words not included in previous dictionary)}} \\
\hline \hline
\multicolumn{5}{||c||}{NEW WORDS} \\ \hline
\multicolumn{1}{||c||}{ANOTHER} & \multicolumn{1}{c||}{COURAGE} & \multicolumn{1}{c||}{HANDLE} & \multicolumn{1}{c||}{MOMENT} & SILENT \\ \hline
\multicolumn{1}{||c||}{BANNER} & \multicolumn{1}{c||}{CULTURE} & \multicolumn{1}{c||}{HEALTH} & \multicolumn{1}{c||}{MOTHER} & SILVER \\ \hline
\multicolumn{1}{||c||}{BEFORE} & \multicolumn{1}{c||}{CUSTOM} & \multicolumn{1}{c||}{HEAVEN} & \multicolumn{1}{c||}{MOTION} & SIMPLE \\ \hline
\multicolumn{1}{||c||}{BENEATH} & \multicolumn{1}{c||}{DANGER} & \multicolumn{1}{c||}{HONEST} & \multicolumn{1}{c||}{NATION} & SPIRIT \\ \hline
\multicolumn{1}{||c||}{BETWEEN} & \multicolumn{1}{c||}{DEEPER} & \multicolumn{1}{c||}{HUNGER} & \multicolumn{1}{c||}{NATURE} & STRENGTH \\ \hline
\multicolumn{1}{||c||}{BLESSED} & \multicolumn{1}{c||}{DESERT} & \multicolumn{1}{c||}{INSIDE} & \multicolumn{1}{c||}{NUMBER} & STRONG \\ \hline
\multicolumn{1}{||c||}{BORDER} & \multicolumn{1}{c||}{DIFFERENT} & \multicolumn{1}{c||}{INSIGHT} & \multicolumn{1}{c||}{OFFBEAT} & THOUGHT \\ \hline
\multicolumn{1}{||c||}{BOTTLE} & \multicolumn{1}{c||}{ECHOES} & \multicolumn{1}{c||}{ISLAND} & \multicolumn{1}{c||}{OPTION} & THROUGH \\ \hline
\multicolumn{1}{||c||}{BREEZE} & \multicolumn{1}{c||}{EMPIRE} & \multicolumn{1}{c||}{JOURNEY} & \multicolumn{1}{c||}{PATIENT} & TRAVEL \\ \hline
\multicolumn{1}{||c||}{BRIDGE} & \multicolumn{1}{c||}{FAMILY} & \multicolumn{1}{c||}{JUSTICE} & \multicolumn{1}{c||}{PEACE} & TUNNEL \\ \hline
\multicolumn{1}{||c||}{BUTTER} & \multicolumn{1}{c||}{FAMOUS} & \multicolumn{1}{c||}{KINDNESS} & \multicolumn{1}{c||}{PEOPLE} & VALUES \\ \hline
\multicolumn{1}{||c||}{BUTTON} & \multicolumn{1}{c||}{FATHER} & \multicolumn{1}{c||}{KINGDOM} & \multicolumn{1}{c||}{POSSIBLE} & VISION \\ \hline
\multicolumn{1}{||c||}{CANDLE} & \multicolumn{1}{c||}{FELLOW} & \multicolumn{1}{c||}{LADDER} & \multicolumn{1}{c||}{REASON} & VOICES \\ \hline
\multicolumn{1}{||c||}{CENTER} & \multicolumn{1}{c||}{FOREST} & \multicolumn{1}{c||}{LANTERN} & \multicolumn{1}{c||}{REMOTE} & WARMTH \\ \hline
\multicolumn{1}{||c||}{CHANGE} & \multicolumn{1}{c||}{FORGIVE} & \multicolumn{1}{c||}{LEGEND} & \multicolumn{1}{c||}{RHYTHM} & WINDOW \\ \hline
\multicolumn{1}{||c||}{CHAPTER} & \multicolumn{1}{c||}{FREEDOM} & \multicolumn{1}{c||}{LIGHTS} & \multicolumn{1}{c||}{SECRET} & WINTER \\ \hline
\multicolumn{1}{||c||}{CHOICE} & \multicolumn{1}{c||}{FROZEN} & \multicolumn{1}{c||}{LITTLE} & \multicolumn{1}{c||}{SERVICE} & WITHOUT \\ \hline
\multicolumn{1}{||c||}{CIRCLE} & \multicolumn{1}{c||}{FUTURE} & \multicolumn{1}{c||}{MARKET} & \multicolumn{1}{c||}{SHADOW} & WONDER \\ \hline
\multicolumn{1}{||c||}{CLOSED} & \multicolumn{1}{c||}{GENTLE} & \multicolumn{1}{c||}{MEMORY} & \multicolumn{1}{c||}{SHELTER} & WONDER \\ \hline
\multicolumn{1}{||c||}{CORNER} & \multicolumn{1}{c||}{GROWTH} & \multicolumn{1}{c||}{MIRACLE} & \multicolumn{1}{c||}{SHELVE} & WORLD \\ \hline \hline
\end{longtable}
\label{tab:Dictionary_simulation3}
\end{small}
}

\restoregeometry

\centering
\newgeometry{top=1cm, bottom=1.5cm, left=0.5cm, right=0.5cm}
\begin{small}
\footnotesize
\begin{longtable}{||p{12cm}||p{3.5cm}||}
\caption{\footnotesize Dictionary used in Simulation 4}  \\
\hline
\hline
\multicolumn{1}{||c||}{\textsc{Sentences}}                                                               & \multicolumn{1}{c||}{\textsc{ERC Topic}}    \\ \hline \hline
DEDICATED TO ADVANCING ACADEMIC KNOWLEDGE   AND ADDRESSING TECHNOLOGICAL CHALLENGES          & Life Sciences                     \\ \hline
DEDICATED TO ADVANCING APPLIED KNOWLEDGE   AND ADDRESSING TECHNOLOGICAL CHALLENGES           & Life Sciences                     \\ \hline
DEDICATED TO ADVANCING BASIC KNOWLEDGE   AND ADDRESSING HEALTH CHALLENGES                    & Social Sciences and Humanities    \\ \hline
DEDICATED TO ADVANCING BIOMEDICAL   KNOWLEDGE AND ADDRESSING SOCIETAL CHALLENGES             & Physical Sciences and Engineering \\ \hline
DEDICATED TO ADVANCING COMPUTATIONAL   CHANGE AND ADDRESSING TECHNOLOGICAL CHALLENGES        & Life Sciences                     \\ \hline
DEDICATED TO ADVANCING COMPUTATIONAL   DIVERSITY AND ADDRESSING TECHNOLOGICAL CHALLENGES     & Life Sciences                     \\ \hline
DEDICATED TO ADVANCING COMPUTATIONAL   IDENTITY AND ADDRESSING TECHNOLOGICAL CHALLENGES      & Life Sciences                     \\ \hline
DEDICATED TO ADVANCING COMPUTATIONAL   KNOWLEDGE AND ADDRESSING TECHNOLOGICAL OUTCOMES       & Life Sciences                     \\ \hline
DEDICATED TO ADVANCING COMPUTATIONAL   KNOWLEDGE AND MANAGING TECHNOLOGICAL CHALLENGES       & Life Sciences                     \\ \hline
DEDICATED TO ADVANCING COMPUTATIONAL   KNOWLEDGE AND OVERCOMING TECHNOLOGICAL CHALLENGES     & Life Sciences                     \\ \hline
DEDICATED TO ADVANCING COMPUTATIONAL   KNOWLEDGE AND SOLVING TECHNOLOGICAL CHALLENGES        & Life Sciences                     \\ \hline
DEDICATED TO ADVANCING COMPUTATIONAL   KNOWLEDGE AND UNDERSTANDING TECHNOLOGICAL CHALLENGES  & Life Sciences                     \\ \hline
DEDICATED TO ADVANCING COMPUTATIONAL   KNOWLEDGE BY ADDRESSING TECHNOLOGICAL CHALLENGES      & Life Sciences                     \\ \hline
DEDICATED TO ADVANCING COMPUTATIONAL   KNOWLEDGE FOR ADDRESSING TECHNOLOGICAL CHALLENGES     & Life Sciences                     \\ \hline
DEDICATED TO ADVANCING COMPUTATIONAL   KNOWLEDGE THROUGH ADDRESSING TECHNOLOGICAL CHALLENGES & Life Sciences                     \\ \hline
DEDICATED TO ADVANCING COMPUTATIONAL   KNOWLEDGE WHILST ADDRESSING TECHNOLOGICAL CHALLENGES  & Life Sciences                     \\ \hline
DEDICATED TO ADVANCING COMPUTATIONAL   LITERACY AND ADDRESSING TECHNOLOGICAL CHALLENGES      & Life Sciences                     \\ \hline
DEDICATED TO ADVANCING COMPUTATIONAL   STUDIES AND ADDRESSING TECHNOLOGICAL CHALLENGES       & Life Sciences                     \\ \hline
DEDICATED TO ADVANCING CULTURAL AWARENESS   AND ADDRESSING SOCIETAL CHALLENGES               & Physical Sciences and Engineering \\ \hline
DEDICATED TO ADVANCING CULTURAL CHANGE   AND ADDRESSING SOCIETAL CHALLENGES                  & Physical Sciences and Engineering \\ \hline
DEDICATED TO ADVANCING CULTURAL   CONSCIOUSNESS AND ADDRESSING SOCIETAL CHALLENGES           & Physical Sciences and Engineering \\ \hline
DEDICATED TO ADVANCING CULTURAL KNOWLEDGE   AND ADDRESSING SOCIETAL CHALLENGES               & Physical Sciences and Engineering \\ \hline
DEDICATED TO ADVANCING CULTURAL KNOWLEDGE   AND ADDRESSING SOCIETAL OUTCOMES                 & Physical Sciences and Engineering \\ \hline
DEDICATED TO ADVANCING CULTURAL KNOWLEDGE   AND ADDRESSING SOCIETAL PROBLEMS                 & Physical Sciences and Engineering \\ \hline
DEDICATED TO ADVANCING CULTURAL KNOWLEDGE   AND ADDRESSING SOCIETAL QUESTIONS                & Physical Sciences and Engineering \\ \hline
DEDICATED TO ADVANCING CULTURAL KNOWLEDGE   AND CONFRONTING SOCIETAL CHALLENGES              & Physical Sciences and Engineering \\ \hline
DEDICATED TO ADVANCING CULTURAL KNOWLEDGE   AND MANAGING SOCIETAL CHALLENGES                 & Physical Sciences and Engineering \\ \hline
DEDICATED TO ADVANCING CULTURAL KNOWLEDGE   AND OVERCOMING SOCIETAL CHALLENGES               & Physical Sciences and Engineering \\ \hline
DEDICATED TO ADVANCING CULTURAL KNOWLEDGE   AND RESOLVING SOCIETAL CHALLENGES                & Physical Sciences and Engineering \\ \hline
DEDICATED TO ADVANCING CULTURAL KNOWLEDGE   IN ADDRESSING SOCIETAL CHALLENGES                & Physical Sciences and Engineering \\ \hline
DEDICATED TO ADVANCING CULTURAL KNOWLEDGE   OR ADDRESSING SOCIETAL CHALLENGES                & Physical Sciences and Engineering \\ \hline
DEDICATED TO ADVANCING CULTURAL KNOWLEDGE   THROUGH ADDRESSING SOCIETAL CHALLENGES           & Physical Sciences and Engineering \\ \hline
DEDICATED TO ADVANCING CULTURAL KNOWLEDGE   WHILST ADDRESSING SOCIETAL CHALLENGES            & Physical Sciences and Engineering \\ \hline
DEDICATED TO ADVANCING CULTURAL STUDIES   AND ADDRESSING SOCIETAL CHALLENGES                 & Physical Sciences and Engineering \\ \hline
DEDICATED TO ADVANCING CULTURAL VALUES   AND ADDRESSING SOCIETAL CHALLENGES                  & Physical Sciences and Engineering \\ \hline
DEDICATED TO ADVANCING CURRENT KNOWLEDGE   AND ADDRESSING HEALTH CHALLENGES                  & Social Sciences and Humanities    \\ \hline
DEDICATED TO ADVANCING CURRENT KNOWLEDGE   AND ADDRESSING SOCIETAL CHALLENGES                & Physical Sciences and Engineering \\ \hline
DEDICATED TO ADVANCING MEDICAL AWARENESS   AND ADDRESSING HEALTH CHALLENGES                  & Social Sciences and Humanities    \\ \hline
DEDICATED TO ADVANCING MEDICAL CHANGE AND   ADDRESSING HEALTH CHALLENGES                     & Social Sciences and Humanities    \\ \hline
DEDICATED TO ADVANCING MEDICAL KNOWLEDGE   AND ADDRESSING HEALTH CHALLENGES                  & Social Sciences and Humanities    \\ \hline
DEDICATED TO ADVANCING MEDICAL KNOWLEDGE   AND ADDRESSING HEALTH OUTCOMES                    & Social Sciences and Humanities    \\ \hline
DEDICATED TO ADVANCING MEDICAL KNOWLEDGE   AND ADDRESSING HEALTH QUESTIONS                   & Social Sciences and Humanities    \\ \hline
DEDICATED TO ADVANCING MEDICAL KNOWLEDGE   AND CHALLENGING HEALTH CHALLENGES                 & Social Sciences and Humanities    \\ \hline
DEDICATED TO ADVANCING MEDICAL KNOWLEDGE   AND CONFRONTING HEALTH CHALLENGES                 & Social Sciences and Humanities    \\ \hline
DEDICATED TO ADVANCING MEDICAL KNOWLEDGE   AND MEETING HEALTH CHALLENGES                     & Social Sciences and Humanities    \\ \hline
DEDICATED TO ADVANCING MEDICAL KNOWLEDGE   AND RESOLVING HEALTH CHALLENGES                   & Social Sciences and Humanities    \\ \hline
DEDICATED TO ADVANCING MEDICAL KNOWLEDGE   AND SOLVING HEALTH CHALLENGES                     & Social Sciences and Humanities    \\ \hline
DEDICATED TO ADVANCING MEDICAL KNOWLEDGE   AND UNDERSTANDING HEALTH CHALLENGES               & Social Sciences and Humanities    \\ \hline
DEDICATED TO ADVANCING MEDICAL KNOWLEDGE   BY ADDRESSING HEALTH CHALLENGES                   & Social Sciences and Humanities    \\ \hline
DEDICATED TO ADVANCING MEDICAL KNOWLEDGE   IN ADDRESSING HEALTH CHALLENGES                   & Social Sciences and Humanities    \\ \hline
DEDICATED TO ADVANCING MEDICAL KNOWLEDGE   OR ADDRESSING HEALTH CHALLENGES                   & Social Sciences and Humanities    \\ \hline
DEDICATED TO ADVANCING MEDICAL KNOWLEDGE   THROUGH ADDRESSING HEALTH CHALLENGES              & Social Sciences and Humanities    \\ \hline
DEDICATED TO ADVANCING MEDICAL KNOWLEDGE   WHILST ADDRESSING HEALTH CHALLENGES               & Social Sciences and Humanities    \\ \hline
DEDICATED TO ADVANCING MEDICAL VALUES AND   ADDRESSING HEALTH CHALLENGES                     & Social Sciences and Humanities    \\ \hline
DEDICATED TO ADVANCING NEW KNOWLEDGE AND   ADDRESSING HEALTH CHALLENGES                      & Social Sciences and Humanities    \\ \hline
DEDICATED TO ADVANCING NEW KNOWLEDGE AND   ADDRESSING SOCIETAL CHALLENGES                    & Physical Sciences and Engineering \\ \hline
DEDICATED TO ADVANCING SCIENTIFIC   ADVANCES AND ADDRESSING HEALTH CHALLENGES                & Social Sciences and Humanities    \\ \hline
DEDICATED TO ADVANCING SCIENTIFIC   ADVANCES AND ADDRESSING TECHNOLOGICAL CHALLENGES         & Life Sciences                     \\ \hline
DEDICATED TO ADVANCING SCIENTIFIC   CONSENSUS AND ADDRESSING SOCIETAL CHALLENGES             & Physical Sciences and Engineering \\ \hline
DEDICATED TO ADVANCING SCIENTIFIC   CONSENSUS AND ADDRESSING TECHNOLOGICAL CHALLENGES        & Life Sciences                     \\ \hline
DEDICATED TO ADVANCING SCIENTIFIC   DISCIPLINES AND ADDRESSING HEALTH CHALLENGES             & Social Sciences and Humanities    \\ \hline
DEDICATED TO ADVANCING SCIENTIFIC   DISCIPLINES AND ADDRESSING TECHNOLOGICAL CHALLENGES      & Life Sciences                     \\ \hline
DEDICATED TO ADVANCING SCIENTIFIC INQUIRY   AND ADDRESSING HEALTH CHALLENGES                 & Social Sciences and Humanities    \\ \hline
DEDICATED TO ADVANCING SCIENTIFIC INQUIRY   AND ADDRESSING SOCIETAL CHALLENGES               & Physical Sciences and Engineering \\ \hline
DEDICATED TO ADVANCING SCIENTIFIC INQUIRY   AND ADDRESSING TECHNOLOGICAL CHALLENGES          & Life Sciences                     \\ \hline
DEDICATED TO ADVANCING SCIENTIFIC   KNOWLEDGE AND ADDRESSING HEALTH CHALLENGES               & Social Sciences and Humanities    \\ \hline
DEDICATED TO ADVANCING SCIENTIFIC   KNOWLEDGE AND ADDRESSING HEALTH GOALS                    & Social Sciences and Humanities    \\ \hline
DEDICATED TO ADVANCING SCIENTIFIC   KNOWLEDGE AND ADDRESSING HEALTH ISSUES                   & Social Sciences and Humanities    \\ \hline
DEDICATED TO ADVANCING SCIENTIFIC   KNOWLEDGE AND ADDRESSING HEALTH OBJECTIVES               & Social Sciences and Humanities    \\ \hline
DEDICATED TO ADVANCING SCIENTIFIC   KNOWLEDGE AND ADDRESSING HEALTH PROBLEMS                 & Social Sciences and Humanities    \\ \hline
DEDICATED TO ADVANCING SCIENTIFIC   KNOWLEDGE AND ADDRESSING SOCIETAL CONCERNS               & Physical Sciences and Engineering \\ \hline
DEDICATED TO ADVANCING SCIENTIFIC   KNOWLEDGE AND ADDRESSING SOCIETAL OBJECTIVES             & Physical Sciences and Engineering \\ \hline
DEDICATED TO ADVANCING SCIENTIFIC   KNOWLEDGE AND ADDRESSING SOCIETAL PROBLEMS               & Physical Sciences and Engineering \\ \hline
DEDICATED TO ADVANCING SCIENTIFIC   KNOWLEDGE AND ADDRESSING SOCIETAL QUESTIONS              & Physical Sciences and Engineering \\ \hline
DEDICATED TO ADVANCING SCIENTIFIC   KNOWLEDGE AND ADDRESSING TECHNOLOGICAL BOUNDARIES        & Life Sciences                     \\ \hline
DEDICATED TO ADVANCING SCIENTIFIC   KNOWLEDGE AND ADDRESSING TECHNOLOGICAL CHALLENGES        & Life Sciences                     \\ \hline
DEDICATED TO ADVANCING SCIENTIFIC   KNOWLEDGE AND ADDRESSING TECHNOLOGICAL CONCERNS          & Life Sciences                     \\ \hline
DEDICATED TO ADVANCING SCIENTIFIC   KNOWLEDGE AND ADDRESSING TECHNOLOGICAL GOALS             & Life Sciences                     \\ \hline
DEDICATED TO ADVANCING SCIENTIFIC   KNOWLEDGE AND ADDRESSING TECHNOLOGICAL ISSUES            & Life Sciences                     \\ \hline
DEDICATED TO ADVANCING SCIENTIFIC   KNOWLEDGE AND ADDRESSING TECHNOLOGICAL OBJECTIVES        & Life Sciences                     \\ \hline
DEDICATED TO ADVANCING SCIENTIFIC   KNOWLEDGE AND ADDRESSING TECHNOLOGICAL PROBLEMS          & Life Sciences                     \\ \hline
DEDICATED TO ADVANCING SCIENTIFIC   KNOWLEDGE AND CHALLENGING SOCIETAL CHALLENGES            & Physical Sciences and Engineering \\ \hline
DEDICATED TO ADVANCING SCIENTIFIC   KNOWLEDGE AND CHALLENGING TECHNOLOGICAL CHALLENGES       & Life Sciences                     \\ \hline
DEDICATED TO ADVANCING SCIENTIFIC   KNOWLEDGE AND CONFRONTING HEALTH CHALLENGES              & Social Sciences and Humanities    \\ \hline
DEDICATED TO ADVANCING SCIENTIFIC   KNOWLEDGE AND CONFRONTING TECHNOLOGICAL CHALLENGES       & Life Sciences                     \\ \hline
DEDICATED TO ADVANCING SCIENTIFIC   KNOWLEDGE AND FACING HEALTH CHALLENGES                   & Social Sciences and Humanities    \\ \hline
DEDICATED TO ADVANCING SCIENTIFIC   KNOWLEDGE AND FACING TECHNOLOGICAL CHALLENGES            & Life Sciences                     \\ \hline
DEDICATED TO ADVANCING SCIENTIFIC   KNOWLEDGE AND MANAGING HEALTH CHALLENGES                 & Social Sciences and Humanities    \\ \hline
DEDICATED TO ADVANCING SCIENTIFIC   KNOWLEDGE AND MANAGING SOCIETAL CHALLENGES               & Physical Sciences and Engineering \\ \hline
DEDICATED TO ADVANCING SCIENTIFIC   KNOWLEDGE AND MANAGING TECHNOLOGICAL CHALLENGES          & Life Sciences                     \\ \hline
DEDICATED TO ADVANCING SCIENTIFIC   KNOWLEDGE AND MEETING HEALTH CHALLENGES                  & Social Sciences and Humanities    \\ \hline
DEDICATED TO ADVANCING SCIENTIFIC   KNOWLEDGE AND OVERCOMING SOCIETAL CHALLENGES             & Physical Sciences and Engineering \\ \hline
DEDICATED TO ADVANCING SCIENTIFIC   KNOWLEDGE AND OVERCOMING TECHNOLOGICAL CHALLENGES        & Life Sciences                     \\ \hline
DEDICATED TO ADVANCING SCIENTIFIC   KNOWLEDGE AND REDUCING HEALTH CHALLENGES                 & Social Sciences and Humanities    \\ \hline
DEDICATED TO ADVANCING SCIENTIFIC   KNOWLEDGE AND REDUCING TECHNOLOGICAL CHALLENGES          & Life Sciences                     \\ \hline
DEDICATED TO ADVANCING SCIENTIFIC   KNOWLEDGE AND SOLVING HEALTH CHALLENGES                  & Social Sciences and Humanities    \\ \hline
DEDICATED TO ADVANCING SCIENTIFIC   KNOWLEDGE AND SOLVING SOCIETAL CHALLENGES                & Physical Sciences and Engineering \\ \hline
DEDICATED TO ADVANCING SCIENTIFIC   KNOWLEDGE AND UNDERSTANDING TECHNOLOGICAL CHALLENGES     & Life Sciences                     \\ \hline
DEDICATED TO ADVANCING SCIENTIFIC   KNOWLEDGE BY ADDRESSING SOCIETAL CHALLENGES              & Physical Sciences and Engineering \\ \hline
DEDICATED TO ADVANCING SCIENTIFIC   KNOWLEDGE FOR ADDRESSING HEALTH CHALLENGES               & Social Sciences and Humanities    \\ \hline \hline
\end{longtable}
\label{tab:Dictionary_simulation4}
\end{small}

\end{document}